\documentclass[showpacs,preprintnumbers,amssymb]{revtex4-2}

\usepackage{graphicx}
\usepackage{dcolumn}
\usepackage[dvipsnames]{xcolor}
\usepackage{bm}
\usepackage{amsmath}
\usepackage{amssymb}
\usepackage{epsfig}
\usepackage{amsfonts}
\usepackage{lineno,hyperref}
\usepackage{array}
\usepackage{float}
\usepackage{microtype}
\usepackage{multirow}
\usepackage{adjustbox}
\usepackage[english]{babel}
\usepackage{epstopdf}
\usepackage{blindtext}
\usepackage{subcaption}
\usepackage[a4paper, total={6.5in, 10in}]{geometry}

\def \a{\alpha}
\def \b{\beta}
\def \l{\lambda}

\def \g{\gamma}

\def \k{\kappa}
\def \o{\omega}

\def \t{\theta}
\def \K{\dot{\phi}^2}
\def \be{\begin{equation}}
\def \ee{\end{equation}}
\def \ben{\begin{eqnarray}}
\def \een{\end{eqnarray}}

\def \G{\bar{G}}
\def \R{\bar{R}}

\def \T{\bar{T}}
\def \p{\bar{p}}
\def \La{\mathcal{L}}
\def \nb{\nabla}

\DeclareUnicodeCharacter{2212}{-}

\begin{document}

\title{NEC violation in $f(\bar{R},\bar{T})$ gravity in the context of a non-canonical theory via modified Raychaudhuri equation }

\author{Arijit Panda}
\email{arijitpanda260195@gmail.com}
\affiliation{Department of Physics, Raiganj University, Raiganj, Uttar Dinajpur, West Bengal, India, 733 134. $\&$\\
Department of Physics, Prabhat Kumar College, Contai, Purba Medinipur, India, 721 404.}

\author{Debashis Gangopadhyay}
\email{debashis.g@snuniv.ac.in}
\affiliation{Department of Physics, School of Natural Sciences, Sister Nivedita University, DG 1/2, Action Area 1 Newtown, Kolkata 700156, India.}

\author{Goutam Manna$^a$}
\email{goutammanna.pkc@gmail.com}
\altaffiliation{$^a$Corresponding author}
\affiliation{Department of Physics, Prabhat Kumar College, Contai, Purba Medinipur 721404, India $\&$\\ Institute of Astronomy Space and Earth Science, Kolkata 700054, India}

\begin{abstract}
In this work, we develop the Raychaudhuri equation in $f(\bar{R},\bar{T})$ gravity in the setting of a non-canonical theory, namely K-essence theory. We solve the modified Raychaudhuri equation for the additive form of $f(\bar{R},\bar{T})$, which is $f_{1}(\bar{R})+f_{2}(\bar{T})$. For this solution, we employ two different scale factors to give two types of $f(\bar{R},\bar{T})$ solutions. The ongoing debate between Fisher et. al. and Harko et. al. in 2020 regarding the additive form of $f(\bar{R},\bar{T})$ may provide a resolution within the modified $f(\bar{R},\bar{T})$ gravity theory. By conducting a viability test and analyzing energy conditions, we have determined that in the first scenario, the null energy condition (NEC) is violated between two regions where the NEC is satisfied. Additionally, we have observed that this violation of the NEC exhibits a symmetric property during the phase transition. These observations indicate that bouncing events may occur as a result of the symmetrical violation of the NEC during the expansion of the universe. Moreover, this model indicates that resonant-type quantum tunneling may take place during the period when the NEC is violated. The findings of NEC violation through the power law of scale factor may have empirical relevance in contemporary observations. In the second scenario, our model indicates that the strong energy condition is violated, but the NEC and weak energy conditions are satisfied. The effective energy density decreases and is positive, while the effective pressure and equation of state parameters are negative. This suggests that the universe is expanding with acceleration and is dominated by dark energy.   
\end{abstract}

\keywords{Cosmology, Modified theories of gravity, Dark energy, Raychaudhuri equation, Friedman equations, K-essence}

\pacs{04.20.-q, 04.20.Cv, 04.50.Kd, 98.80.-k}

\maketitle

\section{Introduction}
In 1955, A. K. Raychaudhuri derived an equation that gives the evolution of a congruence of geodesics observed by other neighbouring congruences of geodesics \cite{Ray52,Ray53,Ray55,Ray57,Ray,Kar}. The author formulated the equation from a purely geometrical standpoint, which is completely independent of any other gravitational theories. The Raychaudhuri equation is typically employed to demonstrate that gravity is attractive in nature, acting between two material objects. Later, it became one of the fundamental lemmas to study the famous Hawking-Penrose singularity theorem \cite{Penrose65,Hawking65,Hawking66,Senovilla,Bhattacharyya}. Over the past twenty years, the Raychaudhuri equation has become important for cosmologists and astrophysicists because of the following aspects : 

(a) The equation considers the non-homogeneity (in terms of shear) and non-isotropy (in terms of rotation) of the universe i.e., on a fundamental note \cite{Kar}.

(b) The equation expresses the rate of expansion of two null or timelike geodesics, which depending upon the energy conditions satisfied by the Ricci tensor may converge or diverge.

(c) The converging geodesics explains the focusing theorem \cite{Poison,Das}, and diverging geodesics explains the non-focusing theorem through the expansion of the universe \cite{Das,Choudhury}.

The Raychaudhuri equation is generally used to study cosmological singularities and has found wide application in the study of various astrophysical phenomena such as gravitational collapse and black hole formation \cite{Choudhury2,Hensh}. For instance, in the context of a collapsing star, the Raychaudhuri equation can be used to analyze several key quantities:

(a) The scalar expansion ($\theta$) is often negative, indicating a decrease in the volume of a small region of the star. As the collapse progresses, $\theta$ becomes more negative, reflecting an increasing rate of contraction.

(b) In cases where the collapsing star is not perfectly spherical or isotropic, significant shear may be present, which is described by the shear tensor ($\sigma_{ab}$). This shear influences the rate of change of the scalar expansion.

(c) The Ricci curvature tensor, which is related to the matter content and energy density, also plays a role. As the star's density increases during collapse, the Ricci tensor changes accordingly, impacting the evolution of $\theta$.

(d) The Raychaudhuri equation is crucial for understanding the conditions under which singularities can form. If $\theta$ becomes increasingly negative and specific criteria, such as the density reaching an infinite value, are met, a singularity may arise where the spacetime curvature becomes infinite.

Thus, the Raychaudhuri equation provides a mathematical framework for understanding the dynamics of expansion and contraction in systems such as a collapsing star under the influence of gravity. It also describes the instability of compact objects due to deviations from equilibrium through the absolute time derivative of expansion \cite{Prisco}. Furthermore, the equation has been employed to study structure formation both within and outside the framework of general relativity (GR), in both weakly and strongly nonlinear regimes \cite{Lobo}, as well as in Palatini $f(R)$ gravity \cite{Bhatti}. Additionally, the Raychaudhuri equation has been utilized in the context of gravitational lensing in astronomy \cite{Kar,Sugiura}.

On the other hand, although it is a powerful theory, the standard General Relativity (GR) has failed to demonstrate all of the cosmological behavior of the universe. The recent results of various observations have confirmed that our universe is expanding rapidly \cite{Riess,Perlmutter,Komatsu}. A compatible theory called  Lambda Cold Dark Matter ($\Lambda$CDM) has been proposed to explain the rapid acceleration of the universe. But it has some serious issues. First, the ``cosmological coincidence problem" \cite{Velten,Yoo2012} asks why the anomalous dark energy component's energy density ($O(meV^4$)) is far lower than what would be predicted based only on quantum field theory. Most dark energy models require extremely fine tuning of their starting energy density, which is orders of 120 smaller than the initial matter-energy density \cite{Yoo2012}. Second, the acceleration of the universe during this specific epoch raises additional questions that the $\Lambda$CDM model is unable to address. A third point that is never emphasized is that, there is profound ambiguity in the definition of $\Lambda$ as the vacuum energy density. Our reason is as follows: It is now accepted folklore that the CDM does not consist of particles as described by the Standard Model of Particle Physics. Therefore, the quantum field theory of such particles and the meaning of vacuum energy for such particles do not exist. Also, the definition of {\it mass} for such unknown particles is highly ambiguous. Standard model particle masses {\it corresponds to poles of the propagator}  as per axiomatic quantum field theory. There does not exist equivalent or other analogues for the particles constituting CDM. Therefore, the so-called vacuum energy density construct is a shot in the dark.  

One approach to answering the above questions is to use modified theories of gravity. In this context,
$f(R)$ gravity \cite{Sotiriou1} and $f(R,T)$ gravity \cite{Harko} theories have been introduced. $f(R)$ gravity generalizes the action of Einstein’s theory by replacing the Ricci scalar ($R$) with a function of $R$. On the other hand, $f(R,T)$ gravity incorporates the trace of energy-momentum tensor ($T$) in the action along with the function of $R$. Besides being a general theory, it has the advantage of producing late-time acceleration without considering any ad hoc term such as $\Lambda$ in the theory. The presence of an additional force operating perpendicular to the test particle's four velocities causes the particle trajectory to diverge from the geodesic path when the stress-energy tensor is taken into account as the source. A source term, corresponding to the matter Lagrangian ($\La_{m}$), is required by the model. Therefore, it follows that different Lagrangians will yield different field equations \cite{Harko}.

Sahoo et al. explored the Bianchi-III and Bianchi$-VI_0$ cosmological models using a string fluid source under $f(R,T)$ gravity, revealing insights into how the universe may have developed under various beginning circumstances and matter distributions\cite{Sahoo2}. Mirza et al. investigated the dynamical system of $f(R,T)$ gravity, taking into account energy conservation, equations of motion, and probable future singularities for a barotropic perfect fluid and a fluid that resembled dark energy. They found no future singularity for the barotropic fluid, but other forms of singularities may occur for dark energy owing to extra degrees of freedom in particular equations of state alternatives. \cite{Mirza}. Naz et al. \cite{Naz} describe the features of anisotropic spherically symmetric star formations under changed $f(R, T)$ gravity. Sharif et al. investigated charged stellar structures with pressure and density connected to a polytropic equation of state in a scenario of $f(R,T)$ gravity \cite{Sharif3}. In \cite{Jasim}, the authors explored the physical characteristics and expected radii of compact stars created by the Tolman-IV complexity-free model in the context of $f(R,T)$ gravity theory. This paper suggests that the complexity strategy used in $f(R,T)$ gravity is a promising way to create astrophysical models that align with current observations. They utilized observational data from the GW190814 event, discovered by the LIGO and Virgo observatories, to test the validity of the Tolman-IV model in $f(R,T)$ gravity. Errehymy et al. \cite{Errehymy} found the simplest solution for traversable wormholes (WHs) using $f(R,T)$ gravity. They used quadratic geometric and linear material modifications to show that the matter composition of WHs may adhere to the energy conditions (ECs). This research considerably extends our knowledge of WHs in the setting of modified gravity, and the results throw fresh light on WH behavior and features, emphasizing their compliance with the $f(R,T)$ gravity framework. Several works have been done in this theory in recent years \cite{Koivisto,Fisher,Harko2,Fisher2,Khoury,Carvalho1,Khoury1,Nojiri8,Carvalho,Ordines,Sharif1,Lobo3,Moraes,Rosa,Rosa2,Fortunato, Maurya,Tangphati}.

The aforementioned modifications to Einstein's equations have been achieved by modifying the left-hand side of Einstein's equations in a non-standard manner, while keeping the right-hand side, which represents the energy-momentum tensor, in its standard or canonical form. In canonical formalism, the Lagrangian of a system can be expressed in the form of $\La=T-V$. However, Raychaudhuri, Goldstein, and Rana, demonstrated that the {\it non-canonical or non-standard} form of the Lagrangian should be the general form \cite{Raychaudhuri,Goldstein,Rana}. Particularly, the non-canonical form of Lagrangian can be reduced to canonical Lagrangian for some specific conditions \cite{Panda}. The {\it K-essence theory} \cite{Das,Vikman,Picon1,Picon2,dg1,dg2,dg3,Chimento,Scherrer,Dutta,Santiago,Mukhanov1,Mukohyama,Babichev1,Visser} is one of the non-canonical theories that explore many possibilities in the study of the universe. The theoretical form of the Lagrangian of the K-essence field is non-canonical in nature which can be expressed as $\La(X,\phi)=-V(\phi)F(X)$ \cite{Vikman,Picon1,Picon2,dg1,dg2,dg3}, or $\La (X,\phi)=F(X)-V(\phi)$ \cite{Dutta,Santiago}, or $\La (X,\phi)\equiv \La (X)= F(X)$ \cite{Scherrer,Mukohyama}, where $F(X)\equiv \La(X) (\neq X)$ is the non-canonical kinetic part, $V(\phi)$ is the canonical potential part and $X=\frac{1}{2} g_{\mu\nu} \nabla^{\mu}\phi \nabla^{\nu} \phi$ is the canonical kinetic part.  This theory emphasizes that the kinetic part dominates over the potential part and demonstrates a dynamic behavior that guarantees late-time acceleration without the need for any fine-tuning. Another intriguing aspect of the K-essence notion is its ability to generate a dark energy component with a sound speed ($c_s$) that is consistently lower than the speed of light. At significant angular scales, this feature might potentially reduce disturbances in the cosmic microwave background (CMB).  Manna et al. \cite{gm1,gm2,gm3,gm4,gm5,gm6,gm7,gm8,gm9} used the Dirac-Born-Infeld (DBI) action \cite{Born,Dirac} to find a key form of the K-essence gravitational metric $\G_{\mu\nu}$. This version of the metric has no explicit conformal equivalency and differs from the standard gravitational metric, $g_{\mu\nu}$. The main difference between K-essence theories with non-canonical kinetic terms and relativistic field theories with canonical kinetic terms is in the dynamical solutions of the K-essence equation of motion. These show both spontaneous Lorentz invariance breaking and the metric changes for the corresponding perturbations. The theoretical expression of the K-essence field Lagrangian demonstrates non-canonical properties, which the metric requires. These properties cause perturbations that spread throughout the emerging curved spacetime, also called analog spacetime. This can mimic the background metric, causing late-time acceleration at the right time in the evolution of the universe. It is important to mention that the Planck collaborations \cite{Planck1,Planck2,Planck3} findings have examined the empirical evidence supporting the concept of K-essence with a DBI-type non-canonical Lagrangian, along with other modified theories. Also, this theory can be used as a unified theory of dark energy and dark matter \cite{Scherrer}. It has also been said that the K-essence theory can be used in a model of dark energy \cite{gm1,gm2,gm6} as well as from a gravitational point of view \cite{gm3,gm4,gm7,gm8} without taking into account the dark parts of the universe. Additionally, it has been revealed that the K-essence theory exhibits a
significant characteristic within the framework of loop quantum cosmology \cite{Chen1,Chen2,Shi}.

Now we would like to mention some works that are relevant to our present work. The researchers in \cite{Choudhury} developed a $f(R)$ gravity model by solving the Raychaudhuri equation and examined the stability of the model. Panda et. al subsequently obtained the functional expression of $f(R,T)$ gravity by using the solutions of the Raychaudhuri equation \cite{Panda1}. However, Das et. al \cite{Das} introduced the Raychaudhuri equation within the context of K-essence geometry. They demonstrated that by imposing a single model, conditionally, it is possible to get three distinct evaluations of congruence in the universe: collapsing, expanding, and static situations. On the other hand, Panda et al. \cite{Panda} have developed the $f(\R,\T)$ gravity model using K-essence geometry. The researchers have formulated the $f(\R,\T)$ gravity theory in this paper by introducing modifications to both sides of the emergent Einstein field equations. The resulting modified field equations are different from the conventional $f(R,T)$ gravity theory proposed by Harko \cite{Harko}, as the underlying geometry is fundamentally distinct. Panda et al. \cite{Panda2} investigated the gravitational collapse in a generalized Vaidya spacetime background using the modified theory of $f(\R,\T)$ gravity. They have indicated that certain selections of $f(\R,\T)$ can lead to the creation of a naked singularity when gravitational collapse occurs. In addition, different selections of $f(\R,\T)$ have been shown to result in a universe that is largely driven by dark energy and is undergoing accelerated expansion.

The objective of the present work is to develop an appropriate version of modified $f(\R,\T)$ gravity by using the solution of the modified Raychaudhuri equation in K-essence and verifying its applicability to contemporary scenarios of the universe.

This paper is organized as follows: Section II briefly summarises the K-essence theory, the $f(\R,\T)$ theory, and the modified Raychaudhuri equation in their corresponding subsections. In Section III, we have constructed the Raychaudhuri equation in $f(\R,\T)$ gravity. Section IV comprises the solutions of the modified Raychaudhuri equation in $f(\R,\T)$ gravity in two different ways, expressed as two different models, described in the subsections. The next section, namely Section V, is about the analysis of two models in two different subsections. In the analysis, we focused on the viability of the models and the energy conditions of these models. The last section, i.e., Section VI, is the conclusion of our work.

\section{Brief review of K-essence, $f(\R,\T)$ gravity and Raychaudhuri equation}

\subsection{The K-essence}

To provide a summary of the K-essence geometry, let's begin with its action \cite{Picon1,Picon2,Babichev1,Vikman}
\ben
S_{k}[\phi,g_{\mu\nu}]= \int d^{4}x {\sqrt -g} \La(X,\phi),
\label{1}
\een
where $X=\frac{1}{2}g^{\mu\nu}\nabla_{\mu}\phi\nabla_{\nu}\phi$ is the canonical kinetic term and $\La(X,\phi)$ is the non-canonical Lagrangian. Here, the standard gravitational metric $g_{\mu\nu}$ has minimal coupling with the K-essence scalar field $\phi$. 

The energy-momentum tensor is defined as \cite{Babichev1,Vikman}:
\ben
T_{\mu\nu}\equiv \frac{-2}{\sqrt {-g}}\frac{\delta S_{k}}{\delta g^{\mu\nu}}=g_{\mu\nu}\La-2\frac{\partial \La}{\partial g^{\mu\nu}}
=g_{\mu\nu}\La+\La_{X}\nabla_{\mu}\phi\nabla_{\nu}\phi,
\label{2}
\een
where $\La_{\mathrm X}= \frac{\partial\La}{\partial X}(\neq 0),~\La_{\mathrm XX}= \frac{\partial^{2}\La}{\partial X^{2}},
~\La_{\mathrm\phi}=\frac{\partial\La}{\partial\phi}$ and $\nabla_{\mu}$ is the covariant derivative defined with respect to the gravitational metric $g_{\mu\nu}$. 

Following \cite{Babichev1,Vikman,gm1,gm2}, we can write a scalar field equation of motion (EOM) as
 
\ben
-\frac{1}{\sqrt {-g}}\frac{\delta S_{k}}{\delta \phi}= \tilde{G}^{\mu\nu}\nabla_{\mu}\nabla_{\nu}\phi +2X\La_{X\phi}-\La_{\phi}=0,
\label{3}
\een
where  
\ben
\tilde{G}^{\mu\nu}\equiv \frac{c_{s}}{\La_{X}^{2}}[\La_{X} g^{\mu\nu} + \La_{XX} \nabla ^{\mu}\phi\nabla^{\nu}\phi]
\label{4}
\een
with $(1+ \frac{2X  \La_{XX}}{L_{X}}) > 0$ and $c_s^{2}(X,\phi)\equiv{(1+2X\frac{\La_{XX}} {\La_{X}})^{-1}}$ \cite{Vikman}.

After a conformal transformation \cite{gm1,gm2,gm3,gm4} $\bar G_{\mu\nu}\equiv \frac{c_{s}}{\La_{X}}G_{\mu\nu}$, we can write the inverse metric of Eq. (\ref{4}) as
\ben
\bar{G}_{\mu\nu}=g_{\mu\nu}-\frac{\La_{XX}}{\La_{X}+2X\La_{XX}}\nabla_{\mu}\phi\nabla_{\nu}\phi.
\label{5}
\een

If the Lagrangian ($\La$) does not directly depend on the K-essence scalar field $\phi$, then EOM (\ref{3}) of the scalar field reduces to
\ben
\frac{1}{\sqrt{-g}}\frac{\delta S_{k}}{\delta \phi}= \bar G^{\mu\nu}\nabla_{\mu}\nabla_{\nu}\phi=0.
\label{6}
\een

The Lagrangian has been considered to be DBI-type \cite{Born,Dirac,Mukohyama,Scherrer,Panda,Panda2} such as
\ben
\La(X,\phi)\equiv \La(X)= 1-\sqrt{1-2X}.
\label{7}
\een
with the condition $\La_{X}\neq 0$ \cite{Vikman}. The form of the metric $\G_{\mu\nu}$ mentioned in Eq. (\ref{5}) has different geometrical structure than the usual $g_{\mu\nu}$. The Lagrangian in Eq. (\ref{7}) exhibits a non-explicit dependence on $\phi$, resulting in modifications to the structural and dynamic characteristics of the solution of the EOM. Therefore, the corresponding K-essence emergent metric (\ref{5}) becomes
\ben
\G_{\mu\nu}=g_{\mu\nu}-\partial_{\mu}\phi\partial_{\nu}\phi
\label{8}
\een
since $\phi$ is a scalar.

The definitions of Christoffel symbol ($\bar{\Gamma}^{\a}_{\mu\nu}$) \cite{gm1,gm2,gm3}, and covariant derivative ($D_{\mu}$) \cite{Babichev1,Vikman,Panda,Panda2} related to K-essence geometry ($\G_{\mu\nu}$) are:
\ben
\bar\Gamma ^{\alpha}_{\mu\nu} 
&&=\Gamma ^{\alpha}_{\mu\nu} -\frac {1}{2(1-2X)}\Big[\delta^{\alpha}_{\mu}\partial_{\nu}
+ \delta^{\alpha}_{\nu}\partial_{\mu}\Big]X~~~~~~~~~~~
\label{9}
\een
and 
\ben
D_{\mu}A_{\nu}=\partial_{\mu} A_{\nu}-\bar \Gamma^{\l}_{\mu\nu}A_{\l}~ \text{or},~
D_{\mu}A^{\nu}=\partial_{\mu} A^{\nu}+\bar \Gamma^{\nu}_{\mu\l}A^{\l}\label{10}
\een
with the inverse emergent metric $\bar G^{\mu\nu}$, such that $\bar G_{\mu\l}\bar G^{\l\nu}=\delta^{\nu}_{\mu}$, where $\Gamma^{\a}_{\mu\nu}$ is the usual Christoffel symbol.

Therefore, ``Emergent Einstein's Field Equation (EEFE)'' reads \cite{Vikman,Panda,Panda2}:
\ben
\bar{\mathcal{G}}_{\mu\nu}=\R_{\mu\nu}-\frac{1}{2}\bar{G}_{\mu\nu}\R=\k \T_{\mu\nu}, \label{11}
\een
where $\k=8\pi G$ is constant, $\R_{\mu\nu}$ is the Ricci tensor and $\R~ (=\R_{\mu\nu}\bar{G}^{\mu\nu})$ is the Ricci scalar and $\T_{\mu\nu}$ is the energy-momentum tensor of the emergent spacetime.

The corresponding definition of $\T_{\mu\nu}$ \cite{Panda,Panda2} is as follows
\ben
\T_{\mu\nu}=-\frac{2}{\sqrt{-\G}}\frac{\partial\Big(\sqrt{-\G}\La(X)\Big)}{\partial \G_{\mu\nu}}
\label{12}
\een
and $\big(-\G \big)^{1/2}=\big(-det({\G_{\mu\nu}})\big)^{1/2}$. It should be noted that the emergent energy-momentum tensor ($\T_{\mu\nu}$) can be calculated from the above definition (\ref{12}) or by solving the left-hand side of EEFE (\ref{11}).

If we consider $g_{\mu\nu}$ to be a flat Friedmann-Lemaître-Robertson-Walker (FLRW) type then the line element of the K-essence emergent metric (\ref{8}) can be written as \cite{gm9,Panda,Panda2}
\ben
dS^{2}=(1-\dot\phi^{2})dt^{2}-a^{2}(t)\sum_{i=1}^{3} (dx^{i})^{2},
\label{13}
\een
where $a(t)$ is the scale parameter and $0<\K<1$ holds for the well-defined behaviour of the line element. It is important to mention that in this context, we utilize a homogeneous K-essence scalar field denoted as $\phi(r,t)\equiv \phi(t)$. Because the dynamical solutions of the K-essence scalar fields cause the spontaneous violation of Lorentz symmetry, the options listed above are the right ones to make.

From the EOM (\ref{6}), we have the relation between the Hubble parameter and the K-essence scalar field as 
\ben
3\frac{\dot a}{a}=3H(t)=-\frac{\ddot\phi}{\dot\phi(1-\dot\phi^{2})},
\label{14}
\een
where $H=\frac{\dot{a}}{a}$ is the Hubble parameter.
This equation leads us to the relation between $\K$ and $a(t)$ as
\ben
a^6=\frac{1-\K}{\K}.
\label{15}
\een
\subsection{$f(\R,\T)$ gravity in the K-essence geometry}
The action in this theory can be written as ($\k=1$)\cite{Panda,Panda2,gm9}
\ben
S=\int d^4x\sqrt{-\G}\Big[f(\R,\T)+\La(X)\Big].
\label{16}
\een

For DBI-type Lagrangian (\ref{7}), the modified field equation is \cite{Panda}
\ben
f_{\R}\R_{\mu\nu}-\frac{1}{2}f(\R,\T)\G_{\mu\nu}+(\G_{\mu\nu}\bar{\square}-D_{\mu}D_{\nu})f_{\R}=\frac{1}{2}\T_{\mu\nu}-f_{\T}\T_{\mu\nu}-f_{\T}\bar{\Theta}_{\mu\nu}.
\label{17}
\een
where $\bar{\square}=D_{\mu}D^{\mu}$, $D_{\mu}$ is the covariant derivative with respect to the metric $\bar{G}_{\mu\nu}$, $f_{\R}=\partial f(\R,\T)/\partial \R$, $f_{\T}=\partial f(\R,\T)/\partial \T$ and 
\ben
\bar{\Theta}_{\mu\nu}=\G^{\alpha\beta}\frac{\partial \T_{\alpha\beta}}{\partial\G^{\mu\nu}}=\G_{\mu\nu}\La(X)-2\T_{\mu\nu}-2\G^{\a\b}\frac{\partial^2 \La(X)}{\partial \G^{\mu\nu}\partial \G^{\a\b}}.
\label{18}
\een

It is worth noting that Panda et al. \cite{Panda} changed both sides of the Einstein field equation using K-essence geometry to obtain the modified field equation \ref{17}). The revised field equation (\ref{17}) differs substantially in geometric characteristics from the equation developed by Harko et al. \cite{Harko}. Also, it is important to mention that the modified field equation (\ref{17}) is not explicitly dependent on the K-essence scalar field or its derivative, but it is implicitly dependent on derivatives of the K-essence scalar field. If we remove all the terms associated with the K-essence scalar field, the modified field equation (\ref{17}) reverts to the precise form derived by Harko et al. \cite{Harko}. It is important to note that Koivisto has already demonstrated in his article \cite{Koivisto} that the conservation equation holds in any modified $f(R,\phi)$ gravity on a purely geometrical background. For $f(\R,\T)$ gravity, it is valid to refer to $\T_{\mu\nu}$ as an energy-momentum tensor only if the condition stated in Eq. (28) of \cite{Panda} is satisfied, which means that the conservation equation in the new geometry $\G_{\mu\nu}$ holds, specifically $D^{\mu}\T_{\mu\nu}=0$. It is interesting to point out that according to the findings in \cite{Carvalho1}, it has been proven that the energy-momentum tensor in any $f(R,T)$ gravity theory is conditionally satisfied. So, the conservation of our energy-momentum tensor ($\T_{\mu\nu}$) \cite{Panda} is satisfied in the K-essence $f(\R,\T)$ gravity theory. The cosmological investigation of this modified field equation (\ref{17}) has been conducted in \cite{Panda}. Taking the Lagrangian mentioned in Eq. (\ref{7}) we can write the Friedmann equations in the K-essence emergent $f(\R,\T)$ gravity as\\

\ben
3H^2=\frac{1}{f_{\R}}\Big[\frac{1-\K}{2}\Big(f(\R,\T)-\R f_{\R}\Big)+\frac{1}{2}\bar{\rho}(1-\K)-\bar{\rho}(1-\K)f_{\T}-f_{\T}\Theta_{00}-3Hf_{\R\R}\dot{\R}\Big]
\label{19}
\een
and
\ben
&&\dot{H}+3H^2(1-\K)=\frac{(1-\K)}{f_{\R}}\Big[-\frac{1}{2}f(\R,\T)+\frac{1}{ (1-\K)}\Big(f_{\R\R\R}\dot{\R}^2+f_{\R\R}\ddot{\R}-\frac{1}{3}\frac{\ddot{\phi}}{\dot{\phi}}f_{\R\R}\dot{\R}\Big)-\nonumber\\
&&\frac{1}{2}\bar{p}+\bar{p}f_{\T}-\frac{f_{\T}}{a^2}\bar{\Theta}_{ii}\Big]
\label{20}
\een

It is important to note that the first Friedmann equation in the standard FLRW cosmology governs the expansion of the universe \cite{Weinberg}. Similarly, in our specific scenario, the modified first Friedmann Eq. (\ref{19}) also governs the expansion of the universe. Following \cite{Panda}, if we consider the equation $3H^2=\tilde{\rho}$, which represents the effective energy density, and the equation $\dot{H}+3H^2(1-\K)=\tilde{p}$, which represents the effective pressure under emergent $f(\R,\T)$ gravity, these two equations has been represented in (\ref{46}) and (\ref{47}) ({\it vide.} Sec.-5.1.3). By combining equations (\ref{46}) and (\ref{47}), we obtain the equation for energy conservation as
\ben
\dot{\tilde{\rho}}+3H (\tilde{\rho}+\tilde{p})=0.
\label{21}
\een

It is noteworthy that the conservation equation is satisfied in this emergent geometry with the effective energy density and effective pressure only. The authors have also shown that Eq. (\ref{19}) and (\ref{20}) can be reduced to the usual gravitational theory of $f(R,T)$ gravity \cite{Panda1} in the absence of the K-essence scalar field ($\phi$) and $\La(X)=\La_{m}$. It should be also mentioned that the authors in \cite{Panda} have achieved an equation of state parameters that satisfies the cosmological observational data for the present universe. This effective energy density ($\tilde{\rho}$) and effective pressure ($\tilde{p}$) in this modified $f(\R,\T)$ gravity can be attributed by the presence of dark energy. Dark energy may be thought as the outcome of the interaction of the K-essence scalar field with the metric. Both $f(R,T)$ gravity and K-essence have already been utilised separately to answer the dark energy theory of the universe \cite{Panda,Cruz,Bhardwaj}.

\subsection{Raychaudhuri equation in K-essence}
In this subsection, we will discuss a brief review of the Raychaudhuri equation in K-essence geometry based on the work in \cite{Das}. For a time-like vector field $v^{\b}$ is connected to an emergent metric tensor $\bar{G}_{\mu\nu}$, the covariant derivative equation (\ref{10}) can be used to write the Raychaudhuri equation in K-essence geometry as \cite{Das},
\ben
\frac{d\tilde{\Theta}}{d\bar{s}}+(D_\a v^\b)(D_\b v^\a)=-\bar{R}_{\g\b}v^\g v^\b
\label{22}
\een
where $\bar{s}$ is an affine parameter and $\tilde{\Theta}\equiv D_\a v^{\b}$ is the scalar expansion of this geometry. 
Following \cite{Das}, we can rewrite the modified Raychaudhuri equation as,

\ben
\frac{d\tilde{\Theta}}{d\bar{s}}+\Big(2{\sigma}^2-2{\omega}^2+\frac{1}{3}\t^2\Big)-\frac{\theta}{(1-2X)}v^\mu\partial_\mu X+\frac{7(v^\mu\partial_\mu X)^{2}}{4{(1-2X)}^2}=-\R_{\gamma\beta}v^\gamma v^\beta
\label{23}
\een
where $\t=\nabla_\a v^\a$ is the usual scalar expansion, symmetric shear $\sigma_{\a\b}=\frac{1}{2}\Big(\nb_\b v_\a+\nb_\a v_\b\Big)-\frac{1}{3}\t h_{\a\b}$,  
anti-symmetric rotation $\o_{\a\b}=\frac{1}{2}\Big(\nb_\b v_\a-\nb_\a v_\b\Big)$, $2\sigma^{2}=\sigma_{\a\b}\sigma^{\a\b}$, $2\o^{2}=\o_{\a\b}\o^{\a\b}$ and 
 $\nb_\b v_\a=\sigma_{\a\b}+\o_{\a\b}+\frac{1}{3}\t h_{\a\b}$ with the three dimensional hypersurface metric $h_{\a\b}=g_{\a\b}-v_\a v_\b$. It is interesting to note that when the K-essence scalar field $\phi$ is not present, Eq. (\ref{23}) can be reduced to the original Raychaudhuri equation in conventional geometry. 

The K-essence scalar field $\phi$ is assumed to be spatially homogeneous and can be chosen as $\phi(x^i,t)=\phi(t)$ \cite{gm3,gm4,gm9}. Based on the results \cite{Das}, we choose to pursue geodesics in a cosmological context using a comoving Lorentz coordinate, $v^{\alpha}=(1,0,0,0)$, the Eq. (\ref{23}) can be written as
\ben
\frac{d\tilde{\Theta}}{d\bar{s}}+\Big(2{\sigma}^2-2{\omega}^2+\frac{1}{3}\t^2\Big)-\frac{\theta~\dot{\phi}~\ddot{\phi}}{(1-\dot{\phi}^2)}+\frac{7\dot{\phi}^2~{\ddot{\phi}}^2}{4{(1-\dot{\phi}^2)}^2}=-\R_{00}.
\label{24}
\een

On the other hand, from the definition of the expansion \cite{Poison,Das} we have 
\ben 
\tilde{\Theta}=D_\a v^\a\equiv\frac{1}{\sqrt{-\G}}~\partial_\a(\sqrt{-\G}~v^\a)=3\frac{\dot{a}}{a}-\frac{\dot\phi\ddot\phi}{(1-\dot\phi^2)},
\label{25}
\een
where $\G=\sqrt{a^6(1-\dot{\phi}^2)}$ is the determinant of the line element (\ref{13}). It should be noted that the usual scalar expansion for considering the flat FLRW spacetime is $\theta=3\frac{\dot{a}}{a}$.
\section{Raychaudhuri equation in the K-essence $f(\R,\T)$ gravity}
The right-hand side of Eqs. (\ref{23}) and (\ref{24}) shows how the Raychaudhuri equation affects gravity using the K-essence geometry through the Ricci tensor ($\R_{\gamma\beta}v^{\gamma}v^{\beta}$). Again Eq. (\ref{17}) can give us the Ricci tensor of the emergent $f(\R,\T)$ gravity, which is 
\ben
\R_{\mu\nu}=\frac{1}{f_{\R}}\Big[\frac{1}{2}\T_{\mu\nu}-f_{\T}\T_{\mu\nu}-f_{\T}\bar{\Theta}_{\mu\nu}+\frac{1}{2}f(\R,\T)-(\G_{\mu\nu}\bar{\square}-D_{\mu}D_{\nu})f_{\R}\Big]
\label{26}
\een

We will now look at the flat emergent K-essence FLRW metric in Eq. (\ref{13}) and the relationship between the scalar field ($\phi$) and the scale parameter ($a(t)$ in Eqs. (\ref{14}) and (\ref{15}). Multiplying both sides of Eq. (\ref{26}) by $v^{\gamma}v^{\beta}$ and assuming that the geodesic to move along comoving frame [$v^{\a}=(1,0,0,0)$] we can write,
\ben
\R_{00}=\frac{1}{f_{\R}}\Big[(\frac{1}{2}-f_{\T})(1-\K)\bar{\rho}-f_{\T}\bar{\Theta}_{00}+\frac{1}{2}
(1-\K)f(\R,\T)-3\frac{\dot{a}}{a}f_{\R\R}\dot{\R}\Big]
\label{27}
\een
$\R_{00}$ and $\bar{\Theta}_{00}$ are the $(00)$ components of $\R_{\mu\nu}$ and $\bar{\Theta}_{\mu\nu}$. In this derivation, we use the perfect fluid energy-momentum tensor ($\T_{\mu\nu}$) in K-essence emergent geometry,
\ben
\T_{\mu}^{\nu}&=& diag(\bar{\rho},-\bar{p},-\bar{p},-\bar{p})=(\bar{\rho} +\bar{p})u_{\mu}u^{\nu}-\delta_{\mu}^{\nu} \bar{p}\nonumber\\
\T_{\mu\nu}&=&\G_{\mu\a}\T^{\a}_{\nu}.
\label{28}
\een

In the above-mentioned context, $\bar{p}$ represents the pressure, whereas $\bar{\rho}$ denotes the energy density of the cosmic fluid in the K-essence emergent gravity. In the co-moving frame, the components of the four-velocity vector are given by $u^{0}=1$ and $u^{i}=0$, where $i$ has values 1, 2, and 3. We can use the ideal fluid model with no vorticity in K-essence theory because of the form of Lagrangian Eq. (\ref{7}), it is not directly dependent on $\phi$. This also shows that the pressure can only be described in terms of the energy density \cite{Vikman, Babichev1}. Under the assumption that the universe is homogeneous and isotropic for large-scale structure, we may select the spatial tensor called the shear to have a value of zero using Frobenius' theorem \cite{Poison,Das}, which means that $2\sigma^2=\sigma^{\a\b}\sigma_{\a\b}=0$. Similarly, the rotation tensor can also be set to zero, represented as $2\o^2=\o^{\a\b}\o_{\a\b}=0$. Now the right-hand side of Eq. (\ref{26}) is governed by the K-essence emergent $f(\R,\T)$ gravity and the rate of change of the expansion ($\frac{d\tilde{\Theta}}{d\bar{s}}$) can be derived from Eq.(\ref{24}). Taking $\theta=3\frac{\dot{a}}{a}$ and equating Eq (\ref{24}) and Eq. (\ref{27}) we compute the {\it modified 
 Raychaudhuri equation in the K-essence $f(\R,\T)$ gravity}, 
\ben
3\frac{\ddot{a}}{a}(1+\K)=-\frac{1}{f_{\R}}\Big[(\frac{1}{2}-f_{\T})(1-\K)\bar{\rho}-f_{\T}\bar{\Theta}_{00}+\frac{1}{2}(1-\K)f(\R,\T)-3H f_{\R\R}\dot{\R}\Big]+g_{1}(\dot{\phi})\nonumber\\
\label{29}
\een
where, $g_{1}$ has been mentioned in (\ref{A1}) ({\it vide.} Appendix A).

The Raychaudhuri equation was derived in the context of the usual $f(R,T)$ gravity by the authors in \cite{Panda1}. Furthermore, it is important to note that the Raychaudhuri equation in the standard $f(R,T)$ gravity was modified while maintaining all the assumptions outlined in the Introduction section. In our specific scenario, the Raychaudhuri equation has been modified to incorporate $f(\R,\T)$ gravity in K-essence geometry, as represented by Eq. (\ref{29}). In this instance, it also maintains all the assumptions that Raychaudhuri took into account. Also, note that the Raychaudhuri equation of $f(\R,\T)$ gravity gives back the Raychaudhuri equation of usual $f(R,T)$ gravity by setting $\K=0$ and $\La(X)=\La_{m}$. It should be noted that Eq. (\ref{29}) has a resemblance to the first Friedmann equation of modified $f(\R,\T)$ gravity (\ref{19}). Given that we are working within the comoving frame, the first Friedmann equation describes the evolution of the universe's expansion in our particular scenario \cite{Weinberg}. In the case where $\K=0$, these two equations become indistinguishable. However, in the context of emergent gravity, there exists a requirement for these two equations to be equal, which is:
\ben
f_{\R}(\dot{H}+H^2)=H^2 \frac{16-49\K}{4}
\label{30}
\een

However, it is important to clarify that our focus in this study is not on the Friedmann equation but rather on the Raychaudhuri equation.

\section{Solutions of the modified Raychaudhuri equation in $f(\R,\T)$ gravity}

To study how the universe works now, we use two types of situations using EOM of K-essence geometry Eqs. (\ref{14}) and (\ref{15}):

\subsection{{\it Case-I:} Considering power law of scale factor}
In this subsection, we use the power law of scale factor and EOM (\ref{14}) to simplify the above Raychaudhuri equation (\ref{29}) in terms of K-essence $f(\R,\T)$ gravity into a second-order differential equation of $f(\R,\T)$ as:

\ben
g_{3}(m,\dot{\phi})\R^{2}f_{\R\R}-g_{2}(m,\dot{\phi})\R f_{\R}+g_{1}(\dot{\phi})f_{\R}-\frac{1}{2}g_{5}(\dot{\phi})f(\R,\T)+g_{5}(\dot{\phi})\bar{\rho}f_{\T}+f_{\T}g_{4}(\dot{\phi})-\frac{1}{2}g_{5}(\dot{\phi})\bar{\rho}=0,
\label{31}
\een

where the expressions of 
$g_{2}(m,\dot{\phi}),~g_{3}(m,\dot{\phi}),~g_{4}(\dot{\phi}),~g_{5}(\dot{\phi})$ have been mentioned in (\ref{A2}), (\ref{A3}), (\ref{A4}), (\ref{A5}) respectively ({\it vide.} Appendix A), and the scale factor is \cite{Kadam,Bahamonde, Cai1}
\ben
a(t)=C \Big(\frac{t}{t_{0}}\Big)^m,~m>0
\label{32}
\een
where $C$ is a constant, $m$ is an arbitrary dimensionless parameter and $t_0$ is a fiducial time. This type of scale factor can be utilized to study the inflationary epoch of the universe \cite{Mukhanov1, Cai1} as well as in the late time era of the universe \cite{Doglov}. In the work of \cite{Cai1, Lima1, Lima2, Lima3} it has been shown observationally that the NEC violation can be occurred in such type of scale factor. The expression of $\K$ in terms of time can be written with the help of EOM of K-essence (\ref{14}) as,
\ben
\K=\frac{1}{1+(C(t/t_{0})^m)^6}.
\label{33}
\een

Now we consider the function $f(\R,\T)$ to be an additive term of $f_{1}(\R)$ and $f_{2}(\T)$ as \cite{Fisher,Panda2,Harko2}
\ben
f(\R,\T)=f_{1}(\R)+f_{2}(\T).
\label{34}
\een

It is important to mention that this specific formulation of $f(\R,\T)$ is entirely valid for deriving meaningful physical quantities, such as $\tilde{\rho}, \tilde{p}$ \cite{Harko2}. Furthermore, the functional expression of $f_{2}(\T)$ carries importance \cite{Carvalho,Ordines}. Exploring the mathematical framework and astrophysical and cosmological consequences of $f(\R,\T)$ gravity is an important field of inquiry that may yield valuable new knowledge. Then Eq. (\ref{31}) can be solved by the method of separation of variables, as $f_{1}(\R)$ and $f_{2}(\T)$ are independent of each other and $f(\R,\T)$ is not an explicit function of $\phi$. In this step, we explain one interesting issue that the form of $f(\R,\T)$ (\ref{34}) has a debate according to Harko et al. \cite{Harko2} and Fisher et al. \cite{Fisher2}. The debate may be resolved as follows: 

In the Ref. \cite{Harko}, the form of $f(R,T)$ is initially expressed as $f_1(R)+f_2(T)$. However, Fisher et al. (2019) \cite{Fisher} have demonstrated that this choice is not novel. They state that the term $f_2(T)$ should be considered as part of the Lagrangian $\La_m$ and therefore does not have any physical significance. However, the correctness of the aforementioned form of $f(R,T)$ was shown in 2020 by Harko et al. \cite{Harko2} after a study published by Fisher. The previous version of $f(R,T)$ is confusing, as demonstrated by Fisher et al. \cite{Fisher2} following Harko's re-establishment. Therefore, in the standard $f(R,T)$ gravity, there is debate on which of the above forms of $f(R,T)$ to use. 

The following, however, may be clarified in terms of K-essence geometry: the form of $f(R,T)$ is not unclear. In the usual $f(R,T)$ gravity theory \cite{Harko}, the energy-momentum tensor is written as 
\begin{equation}
T_{\mu\nu}=g_{\mu\nu}\La_m-2\frac{\partial \La_m}{\partial g_{\mu\nu}}.
    \label{34a}
\end{equation}
For example, considering matter Lagrangian as $\La_m=-p$, $p$ is the pressure of the ideal fluid, the second term of Eq. (\ref{34a}) becomes zero and the energy-momentum tensor, as well as $f_{2}(T)$, becomes the function of $\La_m$ or pressure ($p$) only i.e., $f_{2}(T)\equiv g^{\mu\nu}T_{\mu\nu}=4\La_{m}$. When we put the form of $f_{2}(T)$ in the action, the action becomes 
\begin{eqnarray}
    S=\int d^4x\sqrt{-g}\Big[f(R,T)+\La_{m}\Big]=\int d^4x\sqrt{-g}\Big[f_{1}(R)+\La_{m}^{eff}\Big]
    \label{34b}
\end{eqnarray}
which can be seen in \cite{Fisher, Harko2}. But in K-essence $f(\R,\T)$ gravity theory \cite{Panda},
\begin{eqnarray}
\T_{\mu\nu}
=\G_{\mu\nu}\La(X)-2\frac{\partial \La(X)}{\partial \G^{\mu\nu}},
    \label{34c}
\end{eqnarray}
the second term can never be zero as the Lagrangian is a function of $X(=\frac{1}{2}g^{\mu\nu}\nabla_{\mu}\phi\nabla_{\nu}\phi)$, which cannot be considered simply as the usual pressure. If we consider purely kinetic K-essence with $p=\La(X)$ \cite{Scherrer}, it means the pressure is dependent on the gravitational metric $g^{\mu\nu}$ and first derivative of the K-essence scalar field. In this K-essence geometry, there exist first or second-order derivatives of $p$ as $\frac{\partial p}{\partial X}\equiv \frac{\partial \La(X)}{\partial X}\equiv \La_{X}$ or $\frac{\partial^{2} p}{\partial X^{2}}\equiv \La_{XX}$ terms \cite{Scherrer, Mukhanov1}.   On the other hand, the K-essence emergent gravity metric ($\bar{G}_{\mu\nu}$) (\ref{5}) or (\ref{8}) is not conformally equivalent to the usual gravitational metric ($g_{\mu\nu}$). So, the second term of the above equation (\ref{34c}) cannot be zero. The above explanation is also valid for the second term of the energy-momentum tensor (\ref{2}) of the basic K-essence geometry as $\frac{\partial\La(X)}{\partial g^{\mu\nu}}\neq 0,~\text{since}~\La_{X}\neq 0$ \cite{Mukhanov1, Scherrer}. It is further noted that for the standard K-essence geometry in the study of cosmology, we may employ the energy-momentum tensor (\ref{2}) or (\ref{12}) or (\ref{34c}) as a hydrodynamical fluid, with a redefinition of energy density and pressure (\cite{Mukhanov1}, pp-258). Here, the Lagrangian ($\La(X)$) is not the usual matter Lagrangian ($\La_m$) rather it is a coupling Lagrangian. It is important to mention that in the K-essence geometry, the gravity is minimally coupled with the scalar field $\phi$ i.e., it is an interacting theory (unlike standard scalar field theory).  So there is no question about the absorption of $f_{2}(\T)$ by the Lagrangian ($\La(X)$). Therefore, we may say that the debate for the additive choice of $f(\R,\T)~(=f_{1}(\R)+f_{2}(\T))$ can be resolved through the modified $f(\R,\T)$ gravity theory in the context of K-essence $f(\R,\T)$ gravity theory. Also, note that there exists a non-minimally coupled type K-essence theory such as \cite{Bhattacharya1,Sen}. 

Now, before going further, we want to replace $\bar{\rho}$ in Eq. (\ref{31}) by $\T$ and this can be done by choosing a particular value of the equation of state (EoS) parameter ($\bar{\o}$) defined as, $\bar{\o}=\bar{p}/\bar{\rho}$. We know that for the FLRW universe, the dark energy era has started from $\bar{\o}=-1/3$ \cite{Usmani} and the present value of $\bar{\o}$ is near about $-1$ \cite{Planck1, Planck2}. We chose the EoS parameter to be 
\ben
\bar{\o}=-(\frac{1}{3}+n)
\label{35}
\een
where $n$ is an arbitrary positive constant that can be adjusted to achieve the relationship between $\bar{\rho}$ and $\T$ for different epochs. Eq. (\ref{28}) gives us the trace of the energy-momentum tensor as $\T=\bar{\rho}-3\p$. Our work primarily focuses on the late-time acceleration era and beyond, namely the dark energy era. This is why we consider $\bar{\o}$ as described in Eq. (\ref{35}). With this value of the trace of the energy-momentum tensor, along with the considerations of Eqs. (\ref{32}) and (\ref{33}), we get the solutions of Eq. (\ref{31}) as

\ben
f(\R,\T)=C_{1}\R^{\a}+C_{2}\R^{\gamma}-\frac{2A}{3n}-\frac{\T}{3n}+\Big((\frac{4+6n}{5})B+F\T\Big)^{1+\frac{3n}{2}}C_{3}.
\label{36}
\een

Here, $C_{1},C_{2},~C_{3}$ are unspecified constants of integration. $\a,~\gamma,~B,~F$ are functions that depend on the derivative of $\phi$ and $m$ ({\it vide.} Appendix B). At the same time as we are taking into consideration the form of $f(\R,\T)$ in Eq. (\ref{34}), it is essential to point out that the function $f(\R,\T)$ in the solution (\ref{36}) is implicitly dependent on $\K$, but it is explicitly dependent on $\R$ and $\T$.  It is found that $A$ is the same as the K-essence Lagrangian $\La(X)$ ($=1-\sqrt{1-\K}$), as shown in Eq. (\ref{7}). For $n=\frac{2}{3}~(\bar{\o}=-1)$, we can rewrite the Eq. (\ref{36}) as
\ben
f(\R,\T)=C_{1}\R^{\a}+C_{2}\R^{\g}-\La(X)-\frac{\T}{2}+C_{3}(\frac{8}{5}B+F\T)^{2}.
\label{37}
\een

If we rearrange the above Eq. (\ref{37}), we have

\ben
f(\R,\T)+\La(X)=C_{1}\R^{\a}+C_{2}\R^{\g}-\frac{\T}{2}+C_{3}(\frac{8}{5}B+F\T)^{2}\equiv\mathcal{F}(\R,\T).~~
\label{38}
\een

We observe the following interesting features of the above Eq. (\ref{38}) for a particular value of $\bar{\o}~(=-1)$:

The left side of Eq. (\ref{38}) is the same as when we looked at the total Lagrangian in the action term (\ref{16}). The right side is shown as an additive function of $\R$ and $\T$ with the implicit dependence of $\dot\phi^{2}$, which is the same as when we looked at Eq. (\ref{34}). Therefore, we may conclude that for this specific value of $\bar{\o}$, we return to our initial action (\ref{16}), and it is explicitly dependent on $\R$ and $\T$. In this case, the emerging modified function $\mathcal{F}(\R,\T)$ obscures the coupled Lagrangian $\La(X)$. This modified function $\mathcal{F}(\R,\T)$ plays the role of the total Lagrangian of the system. These can be thought of as the system's emergent phenomena. This specific scenario may be effectively achieved when we are working with higher-order interaction theory. The action (\ref{16}) is like the new $\mathcal{F}(\R,\T)$ gravity theory in this case. The Ricci scalar and the trace of the energy-momentum tensor of the K-essence geometry hide the K-essence scalar field. Here, $\mathcal{F}(\R,\T)$ is explicitly dependent on $\R$ and $\T$ but implicitly dependent on parameters $\dot\phi^{2}$ and $m$. Also, the parameters $\a,~\gamma,~B,~F$ are dependent upon $\dot\phi^{2}$ and $m$, however, they are independent of each other. The dependence of $\dot\phi^2$ comes in Eq. (\ref{38}), since we use $f(\R,\T)$ gravity under the minimally coupled scalar field with the gravity, i.e., K-essence model. Hence, it is not possible to completely remove all the $\dot\phi^2$ terms employing the emergent Ricci scalar and the trace of the energy-momentum tensor. In our scenario, the presence of implicit dependency of the $\dot\phi^2$ terms or interacting terms is consistent. This is a great outcome because if we look at the function (\ref{38}), we choose in the first place to derive the field equation of the $f(\R,\T)$ gravity, i.e., Eq. (\ref{31}) we see that the action gets modified to a theory where the Lagrangian ($\La(X)$) is not present in the action for the particular value of $\bar{\o}~(=-1)$. It should also be noted that the specific value of the EoS parameter ($\bar{\o}=-1$), achieved for the value of $n=2/3$, is same as the standard $\Lambda$CDM model predicts, which is essentially discarded in our particular case. Given that the $\Lambda$CDM model is significantly different from the K-essence geometry, which we employ in our $f(\R,\T)$ gravity computation.

If, on the other hand, any system of the matter Lagrangian ($\La_m$) or the K-essence coupling Lagrangian ($\La(X)$) is zero or hidden by other terms in action, it means that the system is either not consistent in this particular situation or that its dynamics are similar to those of a photon. In this particular scenario, we can say that the equation $2\La=g_{\mu\nu}\frac{dx^{\mu}}{d\lambda}\frac{dx^{\nu}}{d\lambda}=0$ holds, where $\lambda$ represents an affine parameter. One point can also be pointed out that if all integration constants ($C_{1}, C_{2}, C_{3}$) in Eq. (\ref{36}) are set to zero, then the function $f(\R,\T)$ becomes dependent solely on the trace of the energy-momentum tensor ($\T$) and $\K$. This suggests a theory where there is no gravitational geometry component in terms of the Ricci scalar. This situation may be classified as a standard relativistic field theory within the Minkowskian spacetime. It should be noted that in the absence of the K-essence scalar field $\phi$ and $C_{2},~C_{3}$ is taken to be zero, the function $f(\R,\T)$ becomes the usual $f(R,T)$ \cite{Harko} gravity because in the absence of the scalar field, $\R=R$ and $\T=T$. 

\subsection{{\it Case-II:} Considering exponential behaviour of $\K$}
This subsection is devoted to study the $f(\R,\T)$ gravity model via the exponential behavior of $\K$. Maintaining the values of $\K~(0<\K<1)$, we choose the function varies with time as \cite{gm3,gm4,gm7,gm8},
\ben
\K=e^{-t/t_{0}}
\label{39}
\een
where $t_{0}$ is a positive constant to normalize the time scale. With the help of the above Eq. (\ref{39}) and the EOM of K-essence geometry (\ref{14}) we deduce the expression of the scale parameter in terms of $t$ as,
\ben
a(t)=(e^{t/t_{0}}-1)^{1/6}.
\label{40}
\een

Using the above two Eqs. (\ref{39}) and (\ref{40}) in Eq. (\ref{29}), we can derive a new differential form of the Raychaudhuri equation similar to Eq. (\ref{31}) as,
\ben
h_{1}(\dot{\phi})\R^{2}f_{\R\R}-h_{2}(\dot{\phi})\R f_{\R}-\frac{1}{2}g_{5}(\dot{\phi})f(\R,\T)+g_{5}(\dot{\phi})\bar{\rho}f_{\T}+g_{4}(\dot{\phi})f_{\T}-\frac{1}{2}g_{5}(\dot{\phi})\bar{\rho}=0
\label{41}
\een
but different coefficients, which have been mentioned in Eqs. (\ref{A4}), (\ref{A5}), (\ref{A8}) and (\ref{A9}) ({\it vide.} Appendix A). Using the form of $f(\R,\T)$ as Eq. (\ref{34}), and Eq. (\ref{35}) and proceeding similarly as before, we can write the functional form of $f(\R,\T)$ as,

\ben
f(\R,\T)=C'_{1}\R^{\a'}+C'_{2}\R^{\gamma'}-\frac{2A}{3n}-\frac{\T}{3n}+\Big((\frac{4+6n}{5})B'+F'\T\Big)^{1+\frac{3n}{2}}C'_{3}.
\label{42}
\een

The newly formed parameters $(\a',\gamma', B', F')$ has been expressed in Eqs. (\ref{B8}), (\ref{B9}), (\ref{B10}) and (\ref{B11}) ({\it vide.} Appendix B). $C'_{1},~C'_{2}~\&~C'_{3}$ are arbitrary integration constants. 
Like the previous situation, Eq. (\ref{42}) exhibits comparable behavior to Eq. (\ref{36}). When $n$ is equal to $2/3$, this equation exhibits behavior described by Eq.(\ref{38}), where $\bar{\o}$ is equal to $-1$. In the preceding subsection, we have already looked at the behavior of Eq. (\ref{38}) for a specific value of $n$ or $\o$. So, we prioritize studying the theory for a value of $\bar{\o}$ is equal to $-0.9$ for case II.

\begin{figure}
\centering
        \includegraphics[width=0.38\textwidth]{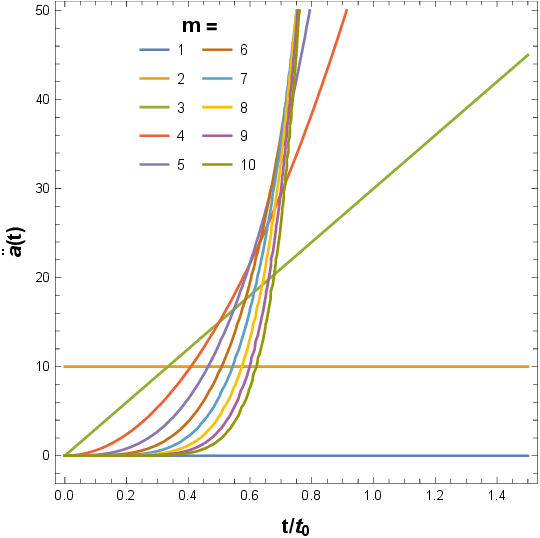}
        \caption{Variation of $\ddot{a}(t)$ with  $t/t_{0}$ for $C=5$ and different $m$ (Case-I)}
        \label{ia}
\end{figure}

\begin{figure}
\centering
        \includegraphics[width=0.38\textwidth]{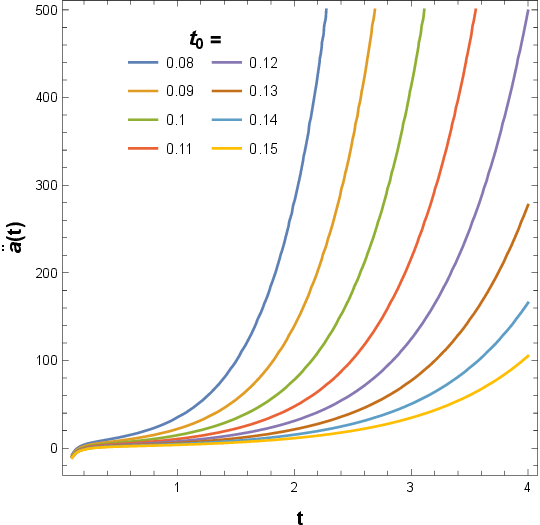}
        \caption{Variation of $\ddot{a}(t)$ with  $t$ for different $t_{0}$ (Case-II)}
        \label{ib}
\end{figure}

Figs. (\ref{ia}) and (\ref{ib}) depict the variations in the $\ddot{a}(t)$ concerning the time parameter, for different values of $m$ and $t_0$, using two types of scale factors (\ref{32}) and (\ref{40}). By observing Fig. (\ref{ia}), it is evident that the sudden increase in the magnitude of $\ddot{a}(t)$ begins approximately around $t/t_{0}=0.2$, whereas in Fig.(\ref{ib}), it initiates at $t=1$. Furthermore, these two pictures illustrate the expanding characteristics of the cosmos. In the next section, we present a model analysis through the viability analysis and energy conditions of our modified Raychaudhuri equation (\ref{29}) in K-essence $f(\R,\T)$ gravity, using the form of $f(\R,\T)$ described in Eqs. (\ref{36}) and (\ref{42}).

\section{Model Analysis}

In this section, we analyze the viability of our model for the two previously described scenarios and examine how it affects the relevant energy conditions, outlined by the Raychaudhuri equations.

\subsection{Model analysis for Case-I}

\subsubsection{The Exponential $\a$ and $\gamma$}
In Eq. (\ref{36}), we see that there exist two exponentials of $\R$ namely $\a$ and $\g$, and they are functions of $m$ and $\K$ (Eqs. (\ref{B1}) and (\ref{B2})). To study the nature of these exponentials, let's plot them concerning $\K$ and $m$. The condition on $\K$ is defined by Eq. (\ref{13}) for the well-behaved nature of spacetime and $m$ is considered to be any positive number (Eq. (\ref{32})). 

\begin{figure*}
\hspace{1cm}
\begin{minipage}[b]{0.37\linewidth}
\centering
 \begin{subfigure}[b]{1.0\textwidth}
        \includegraphics[width=\textwidth]{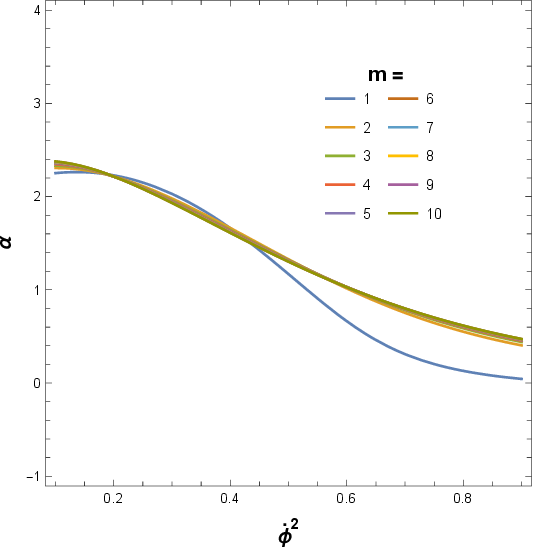}
        \caption{Variation of $\a$ with  $\dot{\phi}^2$ and $m$}
        \label{IIa}
    \end{subfigure}
\end{minipage}
\hspace{2cm}
\begin{minipage}[b]{0.37\linewidth}
\centering
 \begin{subfigure}[b]{1.0\textwidth}
        \includegraphics[width=\textwidth]{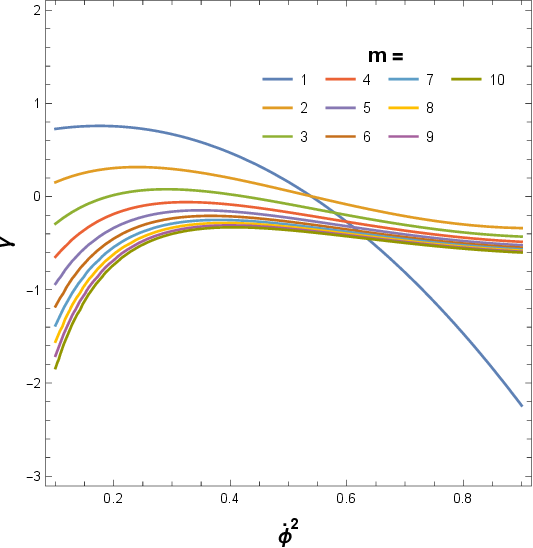}
        \caption{Variation of $\gamma$ with $\dot{\phi}^2$ and $m$}
        \label{IIb}
    \end{subfigure}
\end{minipage}
\caption{Variation of (a) $\a$ and (b) $\gamma$ with $\K$ and $m$.}
\label{II}
\end{figure*}

By referring to Fig. (\ref{IIa}), it is evident that the variable $\a$ is consistently positive. However, Fig. (\ref{IIb}) demonstrates that the variable $\g$ may take negative values for larger values of $m$. The negativity of $\g$ indicates the presence of a singularity in the second term of Eq. (\ref{36}), since the second term approaches infinity as $\R$ approaches zero. In order to eliminate the singularity in the solution (\ref{36}) of $f(\R,\T)$, we choose $C_{2}=0$ throughout our analysis for case I. To ensure simplicity, we have chosen a value of $C_{3}$ as 1 throughout case I. At larger values of $\K$, the quantity $\a$ drops (Fig. \ref{IIa}), indicating a reduction in the exponential of $\R$. This implies that the influence of curvature ($\R$) diminishes at high $\K$. Now, $\K$ may be considered the dark energy density, denoted as $\Omega_{DE}$, measured in units of critical density, related as $\Omega_{Matter} + \Omega_{Radiation} + \Omega_{DE} = 1$ \cite{gm1,gm2,gm6,gm9}. Thus, it may be inferred that as the amount of dark energy increases, the curvature reduces and the attractive effect of gravity decreases. This implies that the objects in the cosmos have less tendency to approach each other, hence contributing to the expansion of the universe.
\subsubsection{Viability Analysis}

The authors in \cite{Sharif,Panda1} pointed out that an additional constraint must be adhered to in $f(R,T)$ gravity, which would also be relevant to our K-essence $f(\R,\T)$ gravity,
\ben
\frac{1+f_{\T}}{f_{\R}}>0.
\label{43}
\een
However, the authors of \cite{Sawicki,Sharif2,Panda1} have provided the viability condition for any $f(R)$ or $f(R,T)$ gravity, which states that $f_{R}(R,T)> 0,~f_{RR}(R,T)>0,~R\geq R_{0},~f_{T}(R,T)>0.$ The comprehensive descriptions of the aforementioned requirements may be found in \cite{Panda1}.  Within the realm of our geometric framework, such entities would correspond to 
\ben
f_{\R}(\R,\T)> 0,~ f_{\R\R}(\R,\T)>0,~ \R\geq \R_{0},~f_{\T}(\R,\T)>0
\label{44}
\een
$\R_0$ is the present Ricci scalar in the context of the K-essence geometry.


\begin{figure*}
\begin{minipage}[t]{0.3\linewidth}
\centering
 \begin{subfigure}[t]{1.0\textwidth}
        \includegraphics[width=\textwidth]{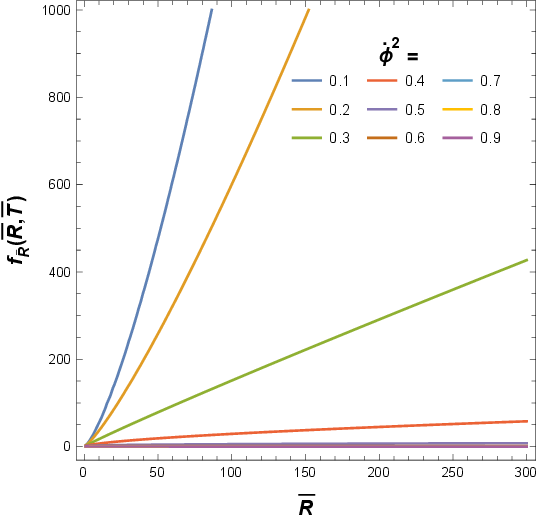}
        \caption{Variation of $f_{\R}(\R,\T)$ with  $\R$ for varying $\K$}
        \label{IIIa}
    \end{subfigure}
\end{minipage}
\hspace{0.1cm}
\begin{minipage}[t]{0.3\linewidth}
\centering
 \begin{subfigure}[t]{1.0\textwidth}
        \includegraphics[width=\textwidth]{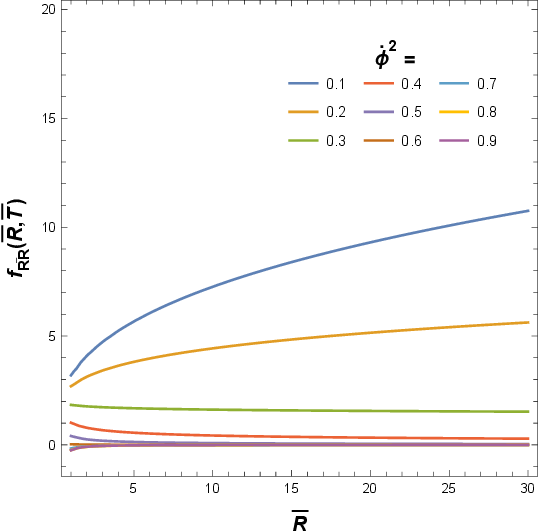}
        \caption{Variation of $f_{\R\R}(\R,\T)$ with $\R$ for varying $\K$}
        \label{IIIb}
    \end{subfigure}
\end{minipage}
\hspace{0.1cm}
\begin{minipage}[t]{0.3\linewidth}
\centering
 \begin{subfigure}[t]{1.0\textwidth}
        \includegraphics[width=\textwidth]{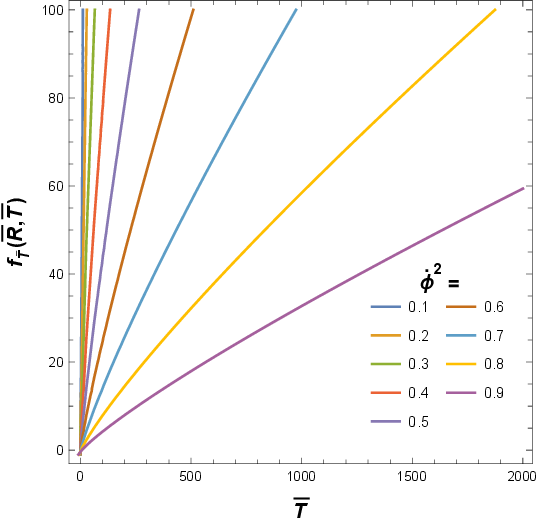}
        \caption{Variation of $f_{\T}(\R,\T)$ with $\T$ for varying $\K$}
        \label{IIIc}
    \end{subfigure}
\end{minipage}
\caption{Viability analysis in terms of $f_{\R}(\R,\T),~f_{\R\R}(\R,\T)~\text{and}~f_{\T}(\R,\T)$ for $m=5$.}
\label{III}
\end{figure*}


\begin{figure*}
\begin{minipage}[t]{0.3\linewidth}
\centering
 \begin{subfigure}[t]{1.0\textwidth}
        \includegraphics[width=\textwidth]{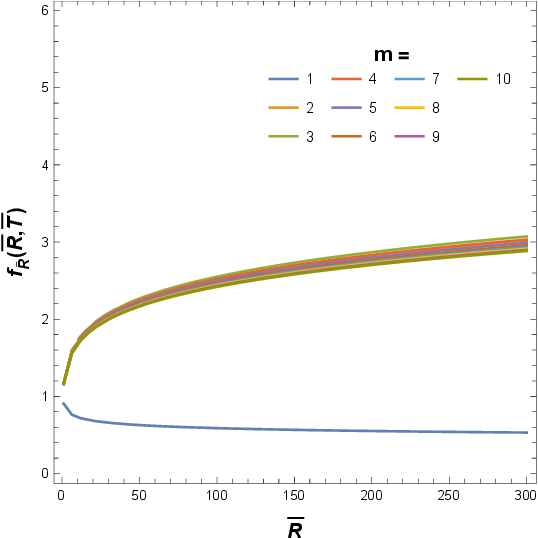}
        \caption{Variation of $f_{\R}(\R,\T)$ with  $\R$ for varying $m$}
        \label{IVa}
    \end{subfigure}
\end{minipage}
\hspace{0.1cm}
\begin{minipage}[t]{0.3\linewidth}
\centering
 \begin{subfigure}[t]{1.0\textwidth}
        \includegraphics[width=\textwidth]{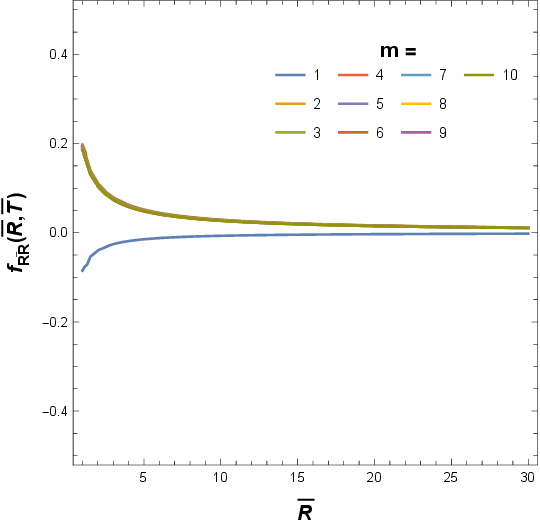}
        \caption{Variation of $f_{\R\R}(\R,\T)$ with  $\R$ for varying $m$}
        \label{IVb}
    \end{subfigure}
\end{minipage}
\hspace{0.1cm}
\begin{minipage}[t]{0.3\linewidth}
\centering
 \begin{subfigure}[t]{1.0\textwidth}
        \includegraphics[width=\textwidth]{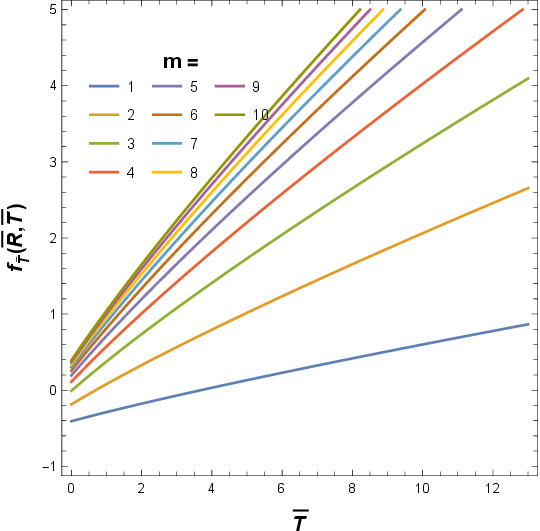}
        \caption{Variation of $f_{\T}(\R,\T)$ with  $\T$ for varying $m$}
        \label{IVc}
    \end{subfigure}
\end{minipage}
\caption{Viability analysis in terms of $f_{\R}(\R,\T),~f_{\R\R}(\R,\T)~\text{and}~f_{\T}(\R,\T)$ for $\K=0.55$.}
\label{IV}
\end{figure*}


Let us analyze the feasibility of the model by examining the above conditions through plots: 
 Fig. (\ref{IIIa}) illustrates the relationship between $f_{\R}$ with respect to $\R$ for $m=5$ and different values of $\K$ ranging from $0$ to $1$. From this figure, it is evident that the function $f_{\R}$ consistently maintains a positive value. At large values of $\K$, the gradient of $f_{\R}$ diminishes, resulting in a nearly flat surface. In other words, at the highest value of dark energy density, the variation in curvature effect becomes almost constant. However, the value of $\K$ can't be negative because its range is limited to $0<\K<1$. Fig. (\ref{IVa}) represents the variation of $f_{\R}$ with $\R$ for varying $m$ and for a constant value of $\K=0.55$. It increases with $\R$ for a higher value of $m$ and remains always positive. These pictures can be found in Tables (\ref{TableA}) and (\ref{TableB}). Therefore, these findings demonstrate the feasibility of our model. Also, note that the positive, nonzero value of $f_{\R}$ is required to validate the requirement (\ref{43}) of any $f(\R,\T)$ gravity theory. 

On the other hand, Fig. (\ref{IIIb}) demonstrates the correlation between $f_{\R\R}$ and $\R$ when $m$ is set to a constant value of $5$ and $\K$ varies. It is non-physical that $\R$ goes to zero. So, we only consider the portion ($\R>0$) for the analysis because of the requirements (\ref{44}). The different variations of $f_{\R\R}$ with $\R$ in Fig. (\ref{IIIb}) are listed in Table (\ref{TableA}). By referring to Table (\ref{TableA}), it is evident that for larger values of $\K$, the region where $f_{\R\R}$ is negative becomes visible. This may be elucidated as follows: If the dark energy density ($\K$) is sufficiently large and substantially dominates the cosmos, an extreme phase occurs. Because a negative value of $f_{\R\R}$ indicates that the system is experiencing a state of maximum conditions. In this situation, the necessary condition ($f_{\R\R}>0$) is not satisfied. Therefore, it may be inferred that when the value of $\K$ is larger, indicating a substantial dominance of dark energy, the cosmos is in an extreme phase. This condition indicates that at the most extreme phase of the cosmos, all the rules of physics can be abundant. These unusual situations may arise as a result of the excessive interactions between the K-essence scalar field and conventional gravity through the new type $f(\R,\T)$ gravity. Currently, the condition $f_{\R\R}<0$ is excluded by ensuring the satisfaction of all the system requirements (\ref{44}). For other values of $\K$, the model is perfectly matched through satisfying requirements (\ref{44}). Fig. (\ref{IVb}) depicts the relationship between $f_{\R\R}$ and $\R$ under the condition that $\K$ remains constant at $0.55$, while various values of $m$ are taken into account. It is feasible over whole areas, except for extremely low values of $m$. From Table (\ref{TableB}), it is evident that for $m=1$, $f_{\R\R}<0$, indicating that the model is inconsistent at this particular value of $m$. This inconsistency is due to the linear variation of the scale factor $a(t)$ with time (\ref{32}). This suggests that the rate of expansion remains constant for $m=1$, but this possibility is eliminated by the current evidence of the universe's rapid expansion. However, almost all portion of the plot  (\ref{IVb}) maintains the condition ($f_{\R\R}>0$).

Figure (\ref{IIIc}) depicts the relationship between $f(\T)$ and $\T$ for various values of $\K$ when $m$ is a constant, whereas Fig. (\ref{IVc}) displays the correlation between $f(\T)$ and $\T$ for different values of $m$ for $\K$ to be a constant. This illustrates that when the parameter $\T$ decreases, the function $f_{\T}$ approaches zero, suggesting that in this particular location, the theory becomes unfeasible while other regions stay feasible. It has been seen that for higher $m(>2)$ value, $f_{\T}$ is always positive (Fig. \ref{IVc}).  It is important to note that the requirements (\ref{44}) must be satisfied throughout the model, ensuring that non-viability is not a concern. Hence, by satisfying the conditions (\ref{44}) of the model, we can conclude that our model is consistent. There are different approaches for determining the stability of this model (such as small perturbation). Considering the compactness of the article we lay up the analysis for the future.

\begin{table}[h]
\begin{center}
\resizebox{8cm}{!}{
\begin{tabular}{|c|c|c|c|} 
\hline
$\R$ & $\K$ & $f_{\R}$ (Fig. \ref{IIIa}) &$f_{\R\R}$ (Fig. \ref{IIIb}) \\
\hline
\multirow{5}{5em}{$0.1$} & $0.1$ & $0.10366$  & $1.40642$ \\

     & $0.3$ & $0.22067$  & $2.08585$ \\

     & $0.5$ & $0.64021$  & $1.99373$ \\

     & $0.61$ & $0.64021$  & $0.03688$ \\ 

     & $0.7$ & $1.27664$  & $-2.63687$ \\

     & $0.9$ & $1.59486$  & $-8.60266$ \\
\hline
\multirow{5}{5em}{$1$} & $0.1$ & $2.35665$ & $3.19714$ \\

     & $0.3$ & $1.94521$  & $1.83865$ \\

     & $0.5$ & $1.31142$  & $0.408407$ \\

      & $0.61$ & $1.00371$  & $0.00372$ \\

    & $0.7$ & $0.79345$  & $-0.16389$ \\

    & $0.9$ & $0.46060$  & $-0.24845$ \\
\hline
\multirow{5}{5em}{$10$} & $0.1$ & $53.57240$  & $7.26789$ \\

    & $0.3$ & $17.14680$  & $1.62074$ \\

    & $0.5$ &  $2.68632$ & $0.08366$ \\

     & $0.61$ &  $1.01231$ & $0.00037$ \\

    & $0.7$ & $0.49314$  & $-0.01019$ \\

    & $0.9$ & $0.13302$  & $-0.00718$ \\
\hline
\end{tabular}}
\end{center}
\caption{Numerical values of $f_{\R}$ (Fig. \ref{IIIa}) and $f_{\R\R}$ (Fig. \ref{IIIb}) for $m=5$ and different values of $\K$ at different curvature($\R$).}
\label{TableA}
 \end{table} 
\begin{table}[h]
\begin{center}
\resizebox{8cm}{!}{
\begin{tabular}{|c|c|c|c|} 
\hline
$\R$ & $m$ & $f_{\R}$ (Fig. \ref{IVa}) &$f_{\R\R}$ (Fig. \ref{IVb})\\
\hline
\multirow{5}{5em}{$0.1$} & $1$ & $1.12461$  & $-1.05444$ \\

     & $3$ & $0.79179$  & $1.34068$ \\

     & $5$ & $0.79651$  & $1.31595$ \\

     & $7$ & $0.79991$  & $1.29796$ \\

     & $9$ & $0.80208$  & $1.28643$ \\
\hline
\multirow{5}{5em}{$1$} & $1$ & $0.90624$ & $-0.08497$ \\

     & $3$ & $1.16932$  & $0.19799$ \\

     & $5$ & $1.16521$  & $0.19251$ \\

    & $7$ & $1.16226$  & $0.18859$ \\

    & $9$ & $1.16039$  & $0.18611$ \\
\hline
\multirow{5}{5em}{$10$} & $1$ & $0.73027$  & $-0.00685$ \\

    & $3$ & $1.72686$  & $0.02924$ \\

    & $5$ &  $1.70459$ & $0.02816$ \\

    & $7$ & $1.68876$  & $0.02740$ \\

    & $9$ & $1.67877$  & $0.02693$ \\
\hline
\end{tabular}}
\end{center}
\caption{Numerical values of $f_{\R}$ (Fig. \ref{IVa}) and $f_{\R\R}$ (Fig. \ref{IVb}) for $\K=0.55$ and different values of $m$ at different curvature($\R$).}
\label{TableB}
 \end{table}

\subsubsection{Energy conditions}
In this subsection, we are going to check the energy conditions for the K-essence $f(\R,\T)$ gravity through the modified Raychaudhuri equation (\ref{29}). To do this, we start with the modified field equation (\ref{17}) of $f(\R,\T)$ gravity, using this we can write the {\it effective energy-momentum tensor} \cite{Panda1} of this geometry as 
\ben
\tilde{T}_{\mu\nu}&&=\R_{\mu\nu}-\frac{1}{2}\G_{\mu\nu}\R \nonumber\\
&&=\frac{1}{f_{\R}}\Big[\frac{1}{2}\T_{\mu\nu}(1-f_{\T})-f_{\T}\bar{\Theta}_{\mu\nu}+\frac{1}{2}f(\R,\T)\G_{\mu\nu}-(\G_{\mu\nu}\bar{\square}-D_{\mu}D_{\nu})f_{\R}-\frac{1}{2}\G_{\mu\nu}\R f_{\R}\Big],
\label{45}
\een
where we consider $\kappa=1$.
Now we consider the matter behaves as a perfect fluid in emergent geometry with energy density $\bar{\rho}$ and pressure $\bar{p}$ as Eq. (\ref{28}) and the geodesics are considered to flow in the co-moving frame ($v_{a}=(1,0,0,0)$). The form of Lagrangian Eq. (\ref{7}) confirms that the pressure can only be described through the energy density and that the perfect fluid model with zero vorticity can be used in K-essence theory \cite{Babichev1,Vikman}. Calculating the $(00)$ and $(ii)$ component of $\bar{T}_{\mu\nu}$ (\ref{28}) and $\bar{\Theta}_{\mu\nu}$ (\ref{18}), we get the expression of effective density ($\tilde{\rho}$) and effective pressure ($\tilde{p}$) in emergent $f(\R,\T)$ gravity as

\ben
\tilde{\rho}=\frac{1}{f_{\R}}\Big[\frac{1}{2}(1-f_{\T})\bar{\rho}(1-\K)+\frac{1}{2}(1-\K)\Big(f(\R,\T)-\R f_{\R}\Big)-f_{\T}\Big(g_{4}(\dot{\phi})-2\bar{\rho}(1-\K)\Big)-3\frac{\dot{a}}{a}f_{\R\R}\dot{\R}\Big]
\label{46}
\een
\ben
\tilde{p}&&=\frac{1}{f_{\R}}\Big[-\frac{1}{2}(1-f_{\T})g_{7}\bar{p}-\frac{1}{2}g_{7}\Big(f(\R,\T)-\R f_{\R}\Big)+\frac{g_{7}}{(1-\K)}\Big(f_{\R\R\R}\dot{\R}^2+f_{\R\R}\ddot{\R}-\frac{1}{3}\frac{\ddot{\phi}}{\dot{\phi}}f_{R\R}\dot{\R}\Big)-\nonumber\\
&&f_{\T}\Big(g_{6}(\dot{\phi})-2\bar{p}g_{7}\Big)\Big].
\label{47}
\een
where, $g_{4}(\dot{\phi})$ (\ref{A4}),~$g_{6} (\dot{\phi})$ (\ref{A6}) and $g_{7}$ (\ref{A7}) ({\it vide.} Appendix A) are functions of $\dot{\phi}$, considered to simplify the calculation. Note that these equations (\ref{46} and \ref{47}) are true for any functional form of $f(\R,\T)$. Following \cite{Panda1}, in this case, the energy conditions can be written in terms of the effective energy density and effective pressure as:\\
{\it (a) Strong energy condition (SEC)}:
\ben
&&\R_{\a\b}v^\a v^\b\geq 0\nonumber\\
&&\Rightarrow \tilde{\rho}+3\tilde{p}\geq 0~ {\text{and}}~\tilde{\rho}+\tilde{p}\geq~0
\label{48}
\een
{\it (b) Null energy condition (NEC)}:
\ben
&&\R_{\a\b}k^\a k^\b\geq~0\nonumber\\
&&\Rightarrow\tilde{\rho}+\tilde{p}\geq 0~
\label{49}
\een
{\it (c) Weak energy condition (WEC)}
\ben
\tilde{\rho}\geq 0~\text{with}~\tilde{\rho}+\tilde{p}\geq 0~
\label{50}
\een
where $v^{\a}$ is a time-like vector field and $k^{\a}$ is a null-like vector field.

To examine the energy conditions mentioned above, it is necessary to develop graphs for different values of certain constants and parameters. The graphs for the energy conditions given below have been plotted to put the functional form of $f(\R,\T)$, derived in Eq. (\ref{36}) in Eqs. (\ref{46}) and (\ref{47}). Given that our study takes into account the homogeneity and isotropy of the cosmos, we may express the effective energy density in terms of the scale factor in solving Eq. (\ref{21}) \cite{Weinberg} as
\ben 
\tilde{\rho}=\tilde{\rho}_{0} a^{-3(1+\tilde{\omega})}
\label{51}
\een
where $\bar{\rho}_{0}=\text{constant}=9.9\times 10^{-30} g/cm^{3}$  being the present energy density according to WMAP data \cite{WMAP} and effective EoS parameter can be defined as $\tilde{\omega}=\frac{\tilde{p}}{\tilde{\rho}}$. Also, note that $\tilde{\omega}$ is not the same as the EoS parameter ($\bar{\omega}$) in K-essence geometry. As we are dealing with the power law expansion of scale factor (\ref{32}), we are interested in investigating the dark energy-dominated late-time era. It should be mentioned that the relation (\ref{51}) is also true for the usual energy density $\bar{\rho}$ as $\bar{\rho}=\bar{\rho}_{0} a^{-3(1+\bar{\omega})}$ in K-essence geometry. We consider the value of $\bar{\omega}$ in this era to be $0<\bar{\omega}\leq-1$ \cite{Planck1}. For an example, we choose $\bar{\omega}=-0.9$, which gives the following relation:
\ben
\bar{p}=-0.9\bar{\rho},~\bar{T}=\bar{\rho}-3\bar{p}=3.7\bar{\rho},~\bar{\rho}=\bar{\rho_{0}}a^{-3/10}.
\label{52}
\een


\begin{figure*}
\begin{minipage}[t]{0.3\linewidth}
\centering
 \begin{subfigure}[t]{1.0\textwidth}
        \includegraphics[width=\textwidth]{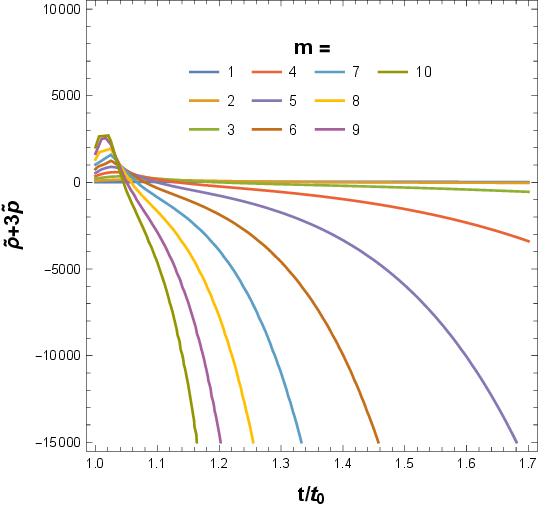}
        \caption{Variation of $\tilde{\rho}+3\tilde{p}$ with $t/t_{0}$ for varying m}
        \label{VIIa}
    \end{subfigure}
\end{minipage}
\hspace{0.1cm}
\begin{minipage}[t]{0.3\linewidth}
\centering
 \begin{subfigure}[t]{1.0\textwidth}
        \includegraphics[width=\textwidth]{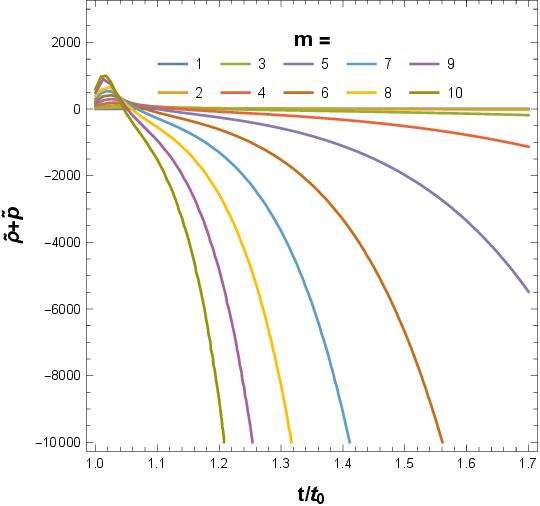}
        \caption{Variation of $\tilde{\rho}+\tilde{p}$ with $t/t_{0}$ for varying m}
        \label{VIIb}
    \end{subfigure}
\end{minipage}
\hspace{0.1cm}
\begin{minipage}[t]{0.3\linewidth}
\centering
 \begin{subfigure}[t]{0.95\textwidth}
        \includegraphics[width=\textwidth]{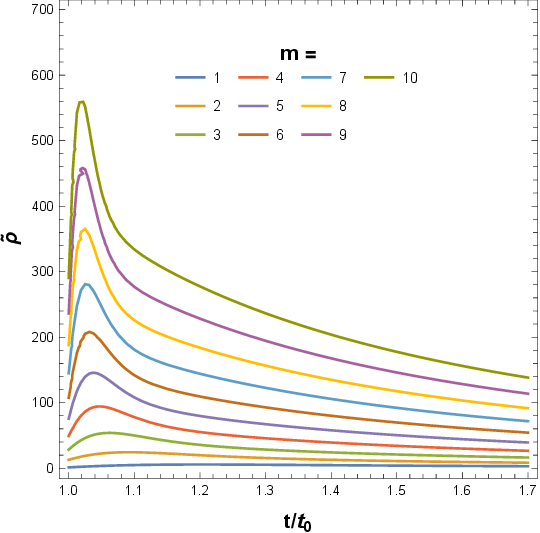}
        \caption{Variation of $\tilde{\rho}$ with $t/t_{0}$ for varying m}
        \label{VIIc}
    \end{subfigure}
\end{minipage}
\caption{Full range of SEC, NEC and WEC with $\bar{\o}=-0.9$ with $C_{1}=1,~C_{2}=0~and~C_{3}=0.1$}
\label{VII}
\end{figure*}
\begin{figure*}
\hspace{1cm}
\begin{minipage}[t]{0.4\linewidth}
\centering
 \begin{subfigure}[t]{1.0\textwidth}
        \includegraphics[width=\textwidth]{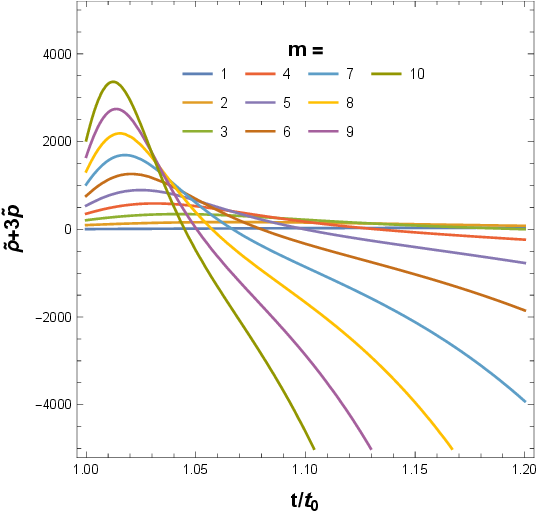}
        \caption{Variation of $\tilde{\rho}+3\tilde{p}$ with $t/t_{0}$ for varying m}
        \label{VIIIa}
    \end{subfigure}
\end{minipage}
\hspace{1cm}
\begin{minipage}[t]{0.4\linewidth}
\centering
 \begin{subfigure}[t]{1.0\textwidth}
        \includegraphics[width=\textwidth]{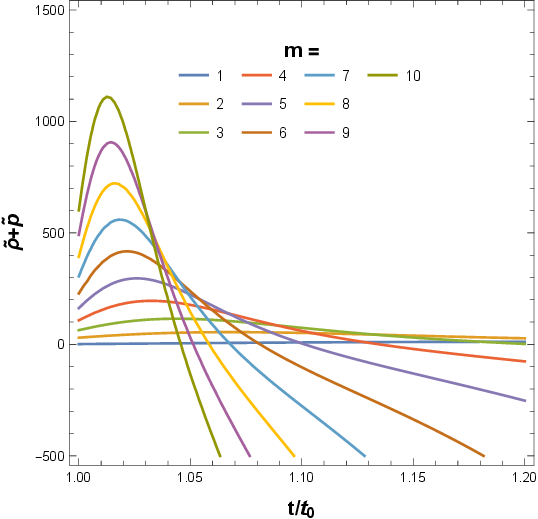}
        \caption{Variation of $\tilde{\rho}+\tilde{p}$ with $t/t_{0}$ for varying m}
        \label{VIIIb}
    \end{subfigure}
\end{minipage}
\caption{Magnified positive range of SEC and NEC with $\bar{\o}=-0.9$ with $C_{1}=1,~C_{2}=0~and~C_{3}=0.1$}
\label{VIII}
\end{figure*}

\begin{table}[h]
\begin{center}
\begin{tabular}{|l|l|l|l|l|} 
\hline
\multicolumn{5}{c}{$\bar{\o}=-0.9$}\\
\hline
 m  & $t/t_{0}$ & $\tilde{\rho}+3\tilde p$ &$\tilde{\rho}+\tilde p$ &  $\tilde{\rho}$ \\
\hline
\multirow{3}{*}{$1$} & $1.0$ & $5.1407$  & $1.3577$   &  $1.2264$ \\

     & $1.3$ & $34.9027$  & $11.6891$ &  $5.2641$\\

     & $1.5$ &  $27.4492$ & $9.1930$ & $3.9765$ \\
\hline 
\multirow{4}{*}{$2$}   & $1.0$ &  $95.6839$ & $29.8663$ & $12.7701$\\ 
  
     & $1.3$ & $33.7174$  & $11.6459$ & $15.5482$ \\ 

      & $1.46$ & $-1.23447$  & $-0.13802$ & $11.5989$ \\ 

     & $1.5$ & $-7.0224$  & $-2.0874$ & $10.9182$\\
\hline 
 \multirow{4}{*}{$3$} & $1.0$ & $207.3671$  & $63.7291$  & $28.3164$ \\

   & $1.21$ & $-11.4486$ &  $-2.54563$ & $34.659$\\
    
    & $1.3$ & $-106.6661$ &   $-34.5738$ & $28.4552$\\

    & $1.5$ & $-98.4147$  &  $$-297.3296$$ & $20.9261$\\
\hline 
\multirow{4}{*}{$5$} & $1.0$ & $539.2576$  & $162.6878$ & $75.6146$\\

& $1.11$ & $-97.0397$ &  $-27.1997$ & $103.091$ \\

      & $1.3$ & $-1740.5449$ & $-577.0452$ & $67.1870$ \\

    & $1.5$ & $-5940.9389$  & $-1977.9633$ & $50.4081$\\
\hline
\multirow{4}{*}{$7$}  & $1.0$ & $1021.3758$ & $305.1387$ &  $144.9242$\\

 & $1.07$ & $-97.6375$ & $-20.6354$ & $211.209$ \\

     & $1.3$ & $-10990.9770$ & $-3656.9788$ & $122.6855$ \\

    & $1.5$ & $-62890.4537$  & $-20958.4675$ & $92.1457$ \\
\hline
\multirow{4}{*}{$9$} & $1.0$ & $1654.7487$  & $491.4250$  & $237.3238$ \\

 & $1.06$ & $-627.091$  & $-188.576$ & $331.052$\\

     & $1.3$ & $-53920.1435$  & $-17961.8014$ & $194.4943$\\

    & $1.5$ &  $-537258.7215$ & $-179077.5432$ & $146.0864$\\
\hline
\end{tabular}
\end{center}
\caption{Energy conditions for $\bar{\o}=-0.9$ (Figs. \ref{VII} and \ref{VIII})}
\label{TableIV}
 \end{table}

\begin{figure*}
\begin{minipage}[t]{0.3\linewidth}
\centering
 \begin{subfigure}[t]{1.0\textwidth}
        \includegraphics[width=\textwidth]{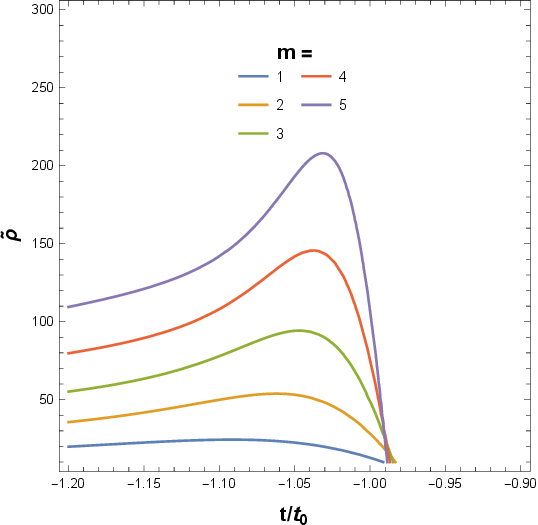}
        \caption{Effective energy density ($\tilde{\rho}$) vs. $t/t_0$: Negative time part}
        \label{IXa}
    \end{subfigure}
\end{minipage}
\hspace{0.1cm}
\begin{minipage}[t]{0.3\linewidth}
\centering
 \begin{subfigure}[t]{1.0\textwidth}
        \includegraphics[width=\textwidth]{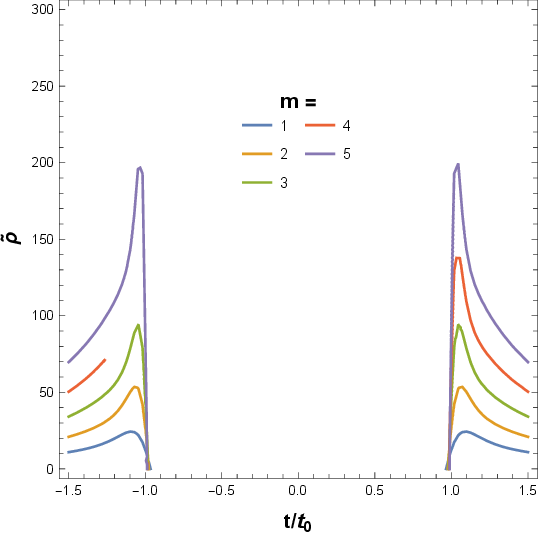}
        \caption{Effective energy density ($\tilde{\rho}$) vs. $t/t_0$: Negative-positive time part}
        \label{IXb}
    \end{subfigure}
\end{minipage}
\hspace{0.1cm}
\begin{minipage}[t]{0.3\linewidth}
\centering
 \begin{subfigure}[t]{0.95\textwidth}
        \includegraphics[width=\textwidth]{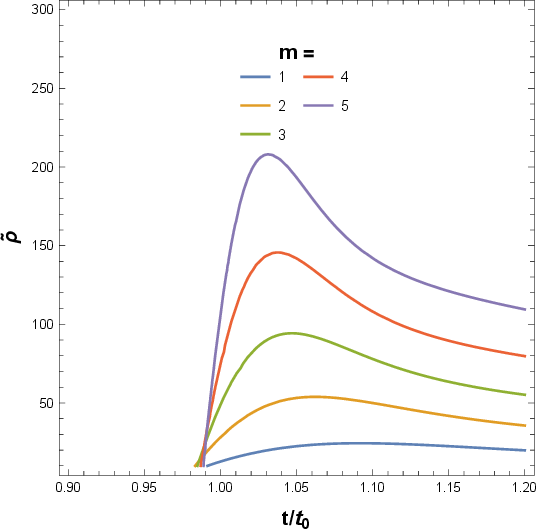}
        \caption{Effective energy density ($\tilde{\rho}$) vs. $t/t_0$: Positive time part  for $\bar{\o}=-0.9$ }
        \label{IXc}
    \end{subfigure}
\end{minipage}
\caption{Effective energy density ($\tilde{\rho}$) vs. $t/t_0$ for $\bar{\o}=-0.9$ }
\label{IX}
\end{figure*}

\begin{figure*}
\begin{minipage}[t]{0.3\linewidth}
\centering
 \begin{subfigure}[t]{1.0\textwidth}
        \includegraphics[width=\textwidth]{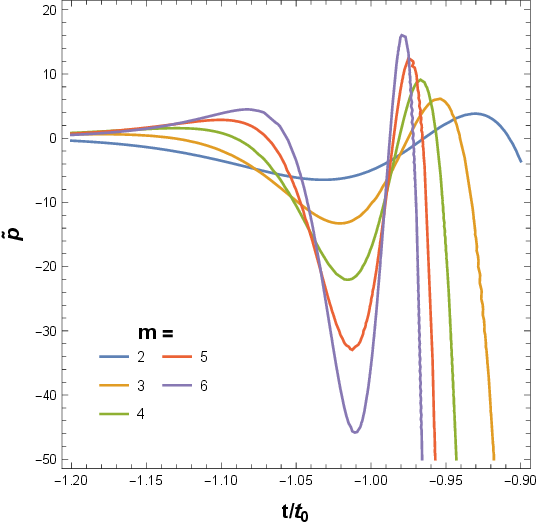}
        \caption{Effective pressure ($\tilde{p}$) vs. $t/t_0$: Negative time part}
        \label{Xa}
    \end{subfigure}
\end{minipage}
\hspace{0.1cm}
\begin{minipage}[t]{0.3\linewidth}
\centering
 \begin{subfigure}[t]{1.0\textwidth}
        \includegraphics[width=\textwidth]{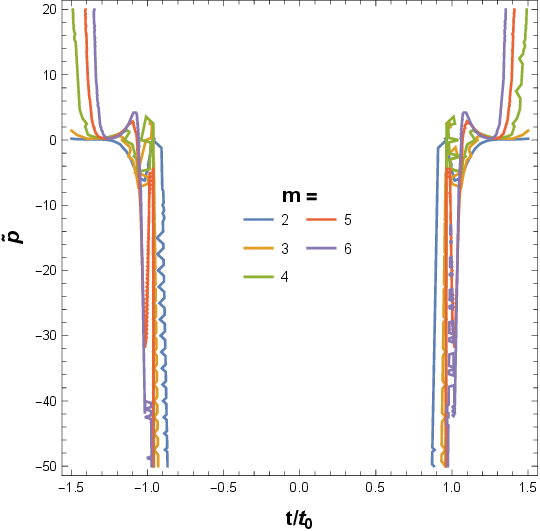}
        \caption{Effective pressure ($\tilde{p}$) vs. $t/t_0$: Negative-positive time part}
        \label{Xb}
    \end{subfigure}
\end{minipage}
\hspace{0.1cm}
\begin{minipage}[t]{0.3\linewidth}
\centering
 \begin{subfigure}[t]{0.95\textwidth}
        \includegraphics[width=\textwidth]{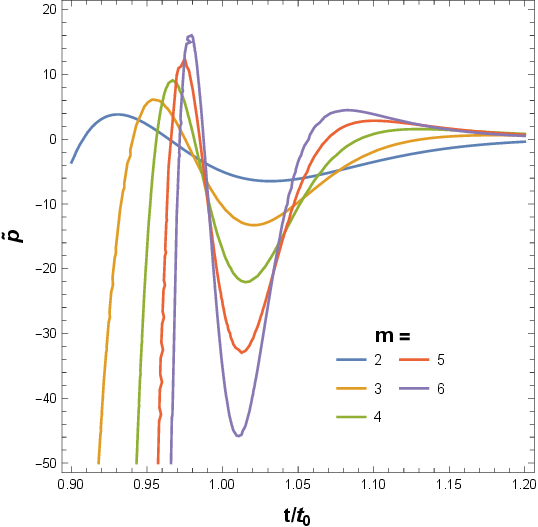}
        \caption{Effective pressure ($\tilde{p}$) vs. $t/t_0$: Positive time part  for $\bar{\o}=-0.9$ }
        \label{Xc}
    \end{subfigure}
\end{minipage}
\caption{Effective pressure ($\tilde{p}$) vs. $t/t_0$ for $\bar{\o}=-0.9$ }
\label{X}
\end{figure*}

\begin{figure}
\centering
\includegraphics[width=0.4\textwidth]{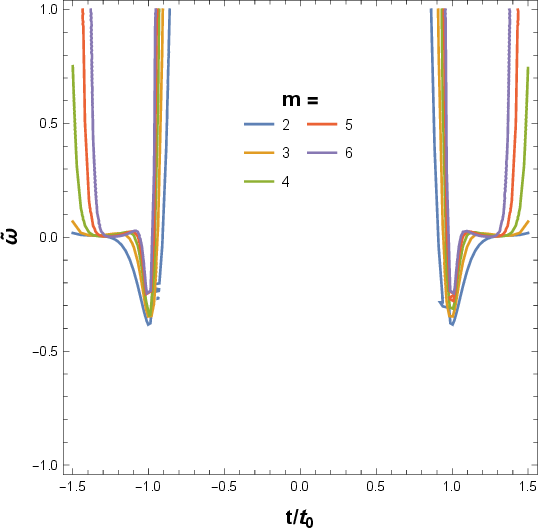}
\caption{Variation of effective EoS parameter $\tilde{\o}$ with $t/t_{0}$ for $\bar{\o}=-0.9$}
\label{XI}
\end{figure}

We consider the constant $C=1$ in Eq. (\ref{32}), and $\K$ is assumed to vary with $a(t)$, as in Eq. (\ref{15}). The following occurrences can arise when $\bar{\o}=-0.9$. Figs. (\ref{VII}) and (\ref{VIII}) together with Table (\ref{TableIV}) demonstrate that the SEC and NEC are violated at $t/t_{0}\geq1.46$ for $m=2$. Fig. (\ref{IX}) and Fig. (\ref{X}) describes the effective energy density ($\tilde{\rho}$) and effective pressure ($\tilde{p}$) respectively for the case of (\ref{36}) in $f(\R,\T)$ gravity, where we consider the EoS  parameter ($\bar{\o}$) to be $-0.9$. Figs. (\ref{IXa}) and (\ref{Xa}) represent the time range ($t/t_{0}$) in the negative direction, whereas Figs. (\ref{IXc}) and (\ref{Xc}) represent the time range ($t/t_{0}$) in the positive direction. In Figs. (\ref{IXb}) and (\ref{Xb}), we present a comprehensive overview of the whole time range ($t/t_{0}$), including both positive and negative values. Fig. (\ref{XI}) represents the variation effective EoS parameter $\tilde{\o}(=\frac{\tilde{p}}{\tilde{\rho}})$ of the form of $f(\R,\T)$ as in Eq. (\ref{36})  with $t/t_{0}$ for K-essence EoS parameter $\bar{\o}=-0.9$.

These phenomena can be described as follows: Table (\ref{TableIV} show that when the range of $t/t_{0}\leq 1.46$ and $m=2$, all the energy criteria are satisfied in our model, even though our major focus is on studying the universe's late-time acceleration. Since we know that if the cosmos expands, SEC must be broken, but other energy criteria such as NEC and WEC have to be satisfied. In this region, $t/t_{0}\leq 1.46$ and $m\leq 2$, our model indicates that the cosmos is not expanding, but can collapse due to satisfied energy requirements. In the context of the conventional Einstein's gravity, it is important to note that when the NEC is satisfied, the energy density in expanding space-time always drops, but in contracting space-time it increases, leading to the eventual collapse of the universe into a singularity. Specifically, it is not feasible to have a bouncing scenario. However, it remains uncertain if this energy criterion may be purposely broken without causing any abnormalities, such as unstable negative-energy states or an imagined speed of sound.

However, if $t/t_0 \geq 1.46$ and the value of $m \geq 2$, we may observe from Table  (\ref{TableIV} that both SEC and NEC are broken, while $\bar{\rho}$ is greater than $0$, and eventually drops from a higher value to lower values with increasing $m$. The violation of the SEC with drops of values in energy density indicates the expanding universe. The violation of the NEC can be explained as follows: The violation of the SEC is evident in the context of an expanding cosmos. However, in our specific example, the NEC is also broken, indicating that our model is describing a bouncing scenario also. It is feasible to achieve a bouncing cosmology with an expanding universe by violating the NEC in modified theories \cite{Arefeva,Arefeva1,Rubakov,Cai,Cai1,Cai2,Singh2023,Ijjas,Sawicki2}. 

The violation of the NEC has enormous consequences in both cosmological and astrophysical settings, with the potential to lead to new insights and interesting physics, as discussed below. In conventional cosmology, the universe's rapid expansion is usually ascribed to dark energy, which is often represented by a cosmological constant. However, NEC violation offers an alternate explanation, implying that the faster expansion may be caused by exotic forms of matter or modified gravity theories that violate the NEC \cite{Caldwell, Novello, Nojiri2004}. This might open up new options for understanding dark energy beyond the cosmological constant model. Furthermore, NEC violation is an essential component in theories of bouncing cosmologies, in which the universe undergoes a contraction phase followed by a bounce leading to expansion, rather than starting from a singularity as in the Big Bang model. In addition, certain inflationary models, particularly those incorporating phantom fields (fields with negative kinetic energy), depend on NEC violations to drive inflation. These models may provide various predictions for the primordial perturbation spectrum, which might then be tested with measurements of the Cosmic Microwave Background (CMB) or large-scale structure. 

Furthermore, it should be noted that in situations when the NEC violates the equation of state parameter $\omega<-1$, resulting in the emergence of what is often referred to as "phantom energy" \cite{Caldwell}.  Various theoretical frameworks, including phantom energy, K-essence and ghost condensates \cite{Hussain}, Horndeski, and beyond-Horndeski theories \cite{Gleyzes}, combine NEC violations to explain cosmic acceleration. Once again, it is important to highlight that observational data plays a crucial role in placing limitations on models that violate the NEC \cite{Riess, Perlmutter}, including the following: The acceleration deduced from the distances of supernovae limits the value of $\omega$. Although $\omega$ values close to $-1$ are preferred, even little variations might indicate violations of the NEC. The anisotropy patterns of the CMB, specifically the Integrated Sachs-Wolfe (ISW) effect, may be used to evaluate models that violate the NEC by providing information on the expansion history of the universe. The expansion rate has a significant impact on the formation of cosmic structures, which offers an additional means of observing and evaluating the validity of models that violate the NEC.

On the other hand, in the astrophysical setting \cite{Morris1988, Cattoen2005, Gimon2009}, we may define it as follows. Traversable wormholes, which are hypothetical tunnels linking distant sections of the universe, need the existence of exotic matters that violate the NEC in order to avoid collapse. Furthermore, NEC violation may be a property of certain unusual stellar objects, such as those made of dark-energy-like fluids or phantom material. These stars would have unexpected features, such as negative pressure or mass, contradicting conventional concepts of stellar structure and development. Also, Violation of the NEC might result in observable naked singularities, possibly refuting this notion and requiring a revision of our understanding of the end states of gravitational collapse.

Now let us discuss some pieces of evidence of NEC violating situations in our universe. In \cite{Singh2023}, authors explored a bounce realization using higher-order curvature in the conventional $f(R, T)$ gravity. The contraction, bounce, and expansion phases were explained by a comprehensive cosmic parameter study. They also find a violation of the null energy requirement, model instability, and a singularity upon deceleration at the bouncing point, suggesting a non-singular bouncing cosmology. Model's ghost condensate behavior at the bouncing point is shown by the equation of state parameter. According to \cite{Arefeva}, field theories that violate the NEC are relevant for solving the cosmic singularity issue and modeling dark energy using the equation of state parameter $\o <-1$. The authors evaluated particle decay rates and discussed mass constraints to protect matter stability in the ghost condensate model. The authors have established a cubic string field theory-derived non-local stringy model with phantom behavior and energy unbounded conditions. The authors also have observed no particle-like excitation and a continuous energy spectrum. In \cite{Arefeva1}, the authors investigated the stability of isotropic cosmological solutions using the Bianchi I model. Applying the result to string field theory models that violate the null energy requirement. The K-essence model and stability of $\phi(t)=t$ solutions were also examined. According to Rubakov \cite{Rubakov}, scalar field theories with second-derivative Lagrangians and second-order field equations were briefly reviewed. Some theories provide solutions that violate the NEC but otherwise appear consistent. They have developed plausible concepts that involve wormhole solutions passing down the throat of the wormhole while violating the NEC. Additionally, they noted that during the future bounce or Genesis period of the cosmos, it would serve as evidence that an occurrence of the NEC occurred in the past. Ijjas et al. \cite{Ijjas} conducted a study on the violation of the NEC in mimetic cosmology. It has been discovered that mimetic cosmology is susceptible to gradient instabilities, even when the NEC is fulfilled. The root cause of the instability lies in the Einstein-Hilbert term present in the action. The matter stress-energy component does not add spatial gradient terms, but instead makes the troublesome curvature modes dynamic. Furthermore, they have demonstrated that mimetic cosmology may be comprehended as a unique boundary of established, orderly theories that incorporate higher-derivative kinetic variables. They also explore methods for eliminating the instability. 

Recently, Cai et al. \cite{Cai,Cai1} noticed a violation of the NEC while investigating inflation, primordial gravity waves, and pulsar timing array measurements. They noted that a primordial NEC violation would cause a blue-tilted GWB spectrum. Its effects on the GWB may be more than designed. They proposed a scenario where the universe has a NEC-violating era following slow-roll inflation with a Hubble parameter of $H\sim H_{inf1}$, followed by a second slow-roll inflation with a larger Hubble parameter of $H (=H_{inf2}>> H_{inf1})$. The primordial gravitational wave spectrum is flat and has a greater amplitude in the cosmic microwave background band, supporting the recent NANOGrav observational results. A Wall-like stochastic GWB spectrum in the right frequency band could be caused by multistage inflation with intermittent NEC violations. They also found stochastic gravitational wave background (SGWB) signals in pulsar timing array (PTA) collaborations, prompting queries into their origins. They again tested a proposed inflationary scenario with an intermediate NEC-violating phase with PTA data. The NEC violation may enhance the primordial tensor power spectrum, providing a possible explanation for the PTA findings. The model matches PTA data in numerical evaluations of NANOGrav's 15-year results. The authors of \cite{Cai2} have investigated the violation of the NEC in cosmology, which has the potential to enhance our comprehension of the primordial universe and its associated gravitational theories. It has been said that the NEC can completely break down in the ``beyond Horndeski" theory, but it is still not clear if this is allowed by the basic features of UV-complete theories or the consistency criteria of effective field theory. The researchers also examined the violation of the NEC at the tree level using perturbative unitarity. They studied this violation in the frameworks of Galilean and ``beyond Horndeski" genesis cosmology, where the cosmos approaches a Minkowskian state in the distant past.

By viewing Figs. (\ref{IX}), (\ref{X}), and (\ref{XI}), it is evident that the effective energy density, pressure, and EoS parameters exhibit a bouncing behavior when the time parameter changes from the negative zone to the positive region. This indicates that the phenomenon of bouncing can occur throughout the transition of the cosmos from the past to the future. Based on the information provided in Fig. (\ref{IX}), it can be observed that the effective energy density consistently drops and eventually reaches zero as the time parameter approaches infinity ($t/t_0 \rightarrow \infty$). This implies that dark energy takes dominance over the matter component of the cosmos. From Fig. (\ref{X}), it can be noted that the effective pressure ($\tilde{p}$) reaches its most minimal negative value during the bouncing time. After a while, it begins to rise and momentarily becomes positive. After that, the value of $\tilde{p}$ remains negative, indicating the universe's accelerated phase. Also, from Fig. (\ref{XI}), it is evident that the effective EoS parameter ($\tilde{\o}$) exhibits a distinct shape around the bouncing point. Specifically, $\tilde{\o}$ does not cross the line at $\tilde{\o}=-1$, which represents the phantom line ($\tilde{\o}=-1$), near the bounce during the subsequent phases. Therefore, we can say that in proximity to the point of bounce, our model does not exhibit behavior like that of a ghost condensate model. These phenomena are similar to studies done by Singh et al. \cite{Singh2023} in the traditional $f(R,T)$ gravity framework and Chakraborty et al. \cite{Chakraborty} in the conventional FLRW cosmology using the Raychaudhuri equation except the behavior of effective parameters ($\tilde{\rho}, \tilde{p}, \tilde{\o}$) in the vicinity of the bouncing point. 

However, in our specific situation, no observable changes are occurring in the vicinity of the temporal interval between $-1$ and $+1$. Before $-1$ and following $+1$, we do see transitions, suggesting the presence of quantum tunneling activity in the transition zone. To clarify, the following explanation can be provided: 

(i) The time region spans from $-1$ to $+1$ and can be compared to the era of NEC violation, as similarly discussed in \cite{Cai,Cai1}. This is followed by a period of slow-roll inflation with a Hubble parameter of approximately $H_{inf1}$. Subsequently, there is a second phase of slow-roll inflation with a larger Hubble parameter, denoted as $H$ (where $H_{inf2} \gg H_{inf1}$). However, it is important to note that our non-canonical DBI type Lagrangian (\ref{7}) does not incorporate any potential dependence. Furthermore, the aforementioned figures suggest the possibility of occurrence inside the gravitational wave spectrum resembling a wall, as indicated in \cite{Cai2}. So we can say that our model case-I indicates that the evolution of the universe goes from one zone that satisfies the NEC to another region that also satisfies the NEC, passing through a region where the NEC is violated, similar to previous cases \cite{Arefeva,Arefeva1,Rubakov,Cai,Cai1,Cai2,Singh2023,Ijjas}.  

(ii) It can be observed from Figs. (\ref{IXb}), (\ref{Xb}), and (\ref{XI}) that a symmetrical pattern may occur on both sides of the NEC violating phase ($-1$ to $+1$). This suggests that when the universe transitions from one region satisfying the NEC to another region satisfying the NEC, passing through the NEC-violating region, it can exhibit symmetry on either side of the NEC-violating phase. Therefore, we may assert that our model indicates the symmetric property while undergoing the phase transition in the NEC-violating phase. Therefore, we can conclude that in our model, bouncing events may occur through a symmetric violation of the NEC throughout the expansion of the cosmos.

(iii) During the time behavior of all the effective parameters mentioned above, resonant-type quantum tunneling events can happen in the time range between $-1$ and $+1$. The quantum behavior seen may arise from the additional interactions between the K-essence scalar field and gravity, in conjunction with higher-dimensional $f(\R,\T)$ gravity. In addition, the rationale for employing a non-canonical DBI-type Lagrangian may be caused by this phenomenon. Furthermore, it may be said that the temporal interval ranging from $-1$ to $+1$ is considered a classically forbidden zone but is acceptable in the realm of quantum theory. This phenomenon can be called the resonant type {\it gravito-quantum} effect in a certain range of time scales. It should be mentioned that the K-essence model exhibits the signature of quantum nature in a different context \cite{gm8}. Also, the article by Bianchi et al. \cite{Bianchi} analyses the phenomenon of quantum tunneling, in which a black hole transitions into a white hole. The study also presents a comprehensive model that describes the whole life cycle of a black hole. The white hole serves as a durable leftover, offering a potential solution to the information dilemma. Therefore, it can be asserted that this type of particular quantum tunneling phenomenon may occur during the transition from a black hole to a white hole within the universe.

Based on the previous discussions of violations of the NEC in the cosmos, it may be inferred that during the late-time acceleration period, a stable violation of the NEC may occur through the bouncing cosmology in the expanding universe. Hence, our model complies with energy requirements during certain intervals of $t/t_{0}$ and $m$, but may violate them in other intervals. These violations of SEC and NEC might arise as a result of our enhanced interactive model utilizing modified gravity ($f(\R, \T)$ gravity) within the framework of a non-canonical theory with minimum coupling between the scalar field and gravity. Also, indications of either quantum tunneling processes or a wall-like GWB spectrum or symmetric property during transitional phases may occur in the bouncing cosmology during the late-time acceleration phase of the universe's expansion history.



\subsection{Model analysis for Case-II}

\subsubsection{The Exponentials $\a'~\text{and}~\gamma'$}
\begin{figure*}
\hspace{0.4cm}
\begin{minipage}[b]{0.38\linewidth}
\centering
 \begin{subfigure}[b]{1.0\textwidth}
        \includegraphics[width=\textwidth]{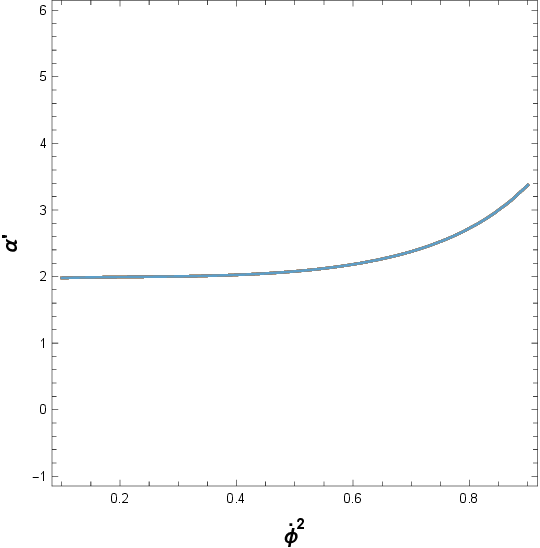}
        \caption{Variation of $\a'$ with $\K$}
        \label{XIIa}
    \end{subfigure}
\end{minipage}
\hspace{1.5cm}
\begin{minipage}[b]{0.38\linewidth}
\centering
 \begin{subfigure}[b]{1.0\textwidth}
        \includegraphics[width=\textwidth]{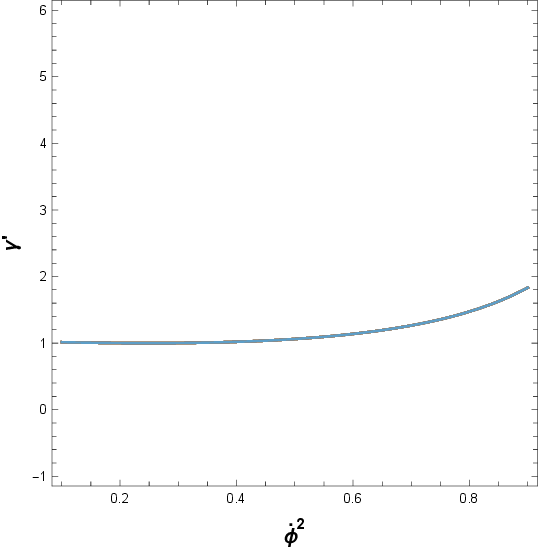}
        \caption{Variation of $\gamma '$ with $\K$}
        \label{XIIb}
    \end{subfigure}
\end{minipage}
\caption{Variation of $\a'$ and $\gamma'$ with $\K$}
\label{XII}
\end{figure*}

Unlike the first case (Case-I) both the factors $\a'$ and $\gamma'$ in Eq. (\ref{42}) shows positive nature with $\K$ (Fig. \ref{XII}). Therefore, in the analysis of this function (\ref{42}) we choose $C'_{1}=1,~C'_{2}=1~\text{and}~C'_{3}=0.1$ everywhere. Note that, the exponentials (\ref{B8} and \ref{B9}) in this case are only functions of $\K$. Also, the range of $\K$ has been chosen to lie between $0.1$ and $0.9$, which is a necessary condition for the well-behaved K-essence FLRW metric (\ref{13}). From Fig. (\ref{XIIa}), we see that the exponential $\a'$ starts at a value of $2$ when $\K \rightarrow 0.1$. Then it increases up to a little bit higher value than $3$ at $\K\rightarrow 0.9$. On the other hand, from Fig. (\ref{XIIb}), $\gamma'$ starts at a value of $2$ (at $\K\rightarrow 0.1$) and ends up being around $2$ (at $\K\rightarrow 0.9$). It is clear from Eq. (\ref{42}) that, as the value of $\K$ increases, which can be considered the dark energy density as already mentioned in the previous case, the exponentials ($\a'$ and $\gamma'$) increase, resulting in a higher value of curvature ($\R$).

\subsubsection{Viability Analysis}
\begin{figure*}
\begin{minipage}[b]{0.3\linewidth}
\centering
 \begin{subfigure}[b]{1.0\textwidth}
        \includegraphics[width=\textwidth]{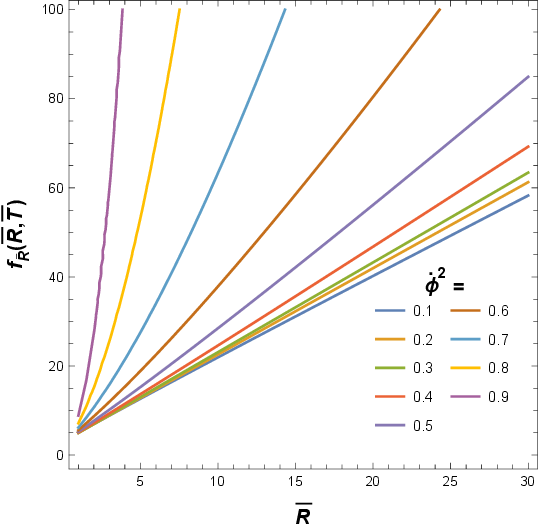}
        \caption{Variation of $f_{\R}(\R,\T)$ with $\R$}
        \label{XIIIa}
    \end{subfigure}
\end{minipage}
\hspace{0.1cm}
\begin{minipage}[b]{0.3\linewidth}
\centering
 \begin{subfigure}[b]{1.0\textwidth}
        \includegraphics[width=\textwidth]{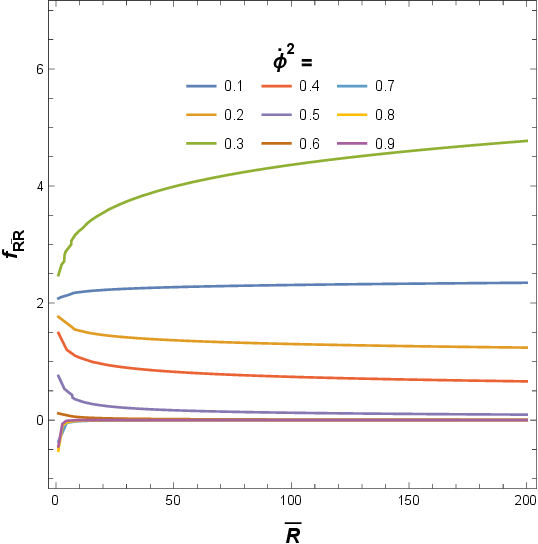}
        \caption{Variation of $f_{\R\R}(\R,\T)$ with $\R$}
        \label{XIIIb}
    \end{subfigure}
\end{minipage}
\hspace{0.1cm}
\begin{minipage}[b]{0.3\linewidth}
\centering
 \begin{subfigure}[b]{1.0\textwidth}
        \includegraphics[width=\textwidth]{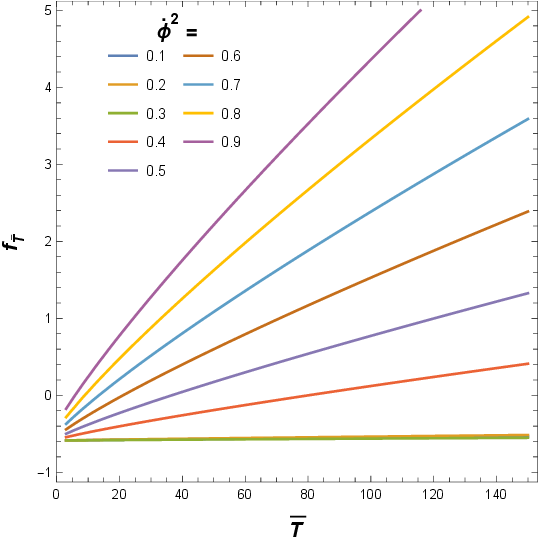}
        \caption{Variation of $f_{\T}(\R,\T)$ with $\T$}
        \label{XIIIc}
    \end{subfigure}
\end{minipage}
\caption{Variation of $f_{\R},f_{\R\R}$ and $f_{\T}$ for different $\K$}
\label{XIII}
\end{figure*}

We recall the viability conditions for $f(\R,\T)$ gravity as mentioned in Eq. (\ref{43}) and Eq. (\ref{44}). We plot the function $f_{\R}(\R,\T)$ in Fig.(\ref{XIIIa}) with $\R$ for different values of $\K$. The function $f_{\R\R}(\R,\T)$ has been plotted in Fig. (\ref{XIIIb}) with $\R$ and $f_{\T}(\R,\T)$ has been plotted in Fig. (\ref{XIIIc}) with $\T$.  Both the graphs (\ref{XIIIa}) and (\ref{XIIIb}) show similar nature as in the first case. But $f_{\T}$ in Fig. (\ref{XIIIc}) stays positive at higher $\T$ and only for the values of $\K$ which are greater than $0.4$. So, the viability conditions holds for high values of $\K$ and $\T$. The non-viability at low $\K$ and $\T$ value may be caused due to interactions of gravity with the K-essence scalar field through the trace of the energy-momentum tensor of this geometry.

\subsubsection{Energy Conditions}
\begin{figure*}
\begin{minipage}[t]{0.3\linewidth}
\centering
\begin{subfigure}[t]{1.0\textwidth}
\includegraphics[width=\textwidth]{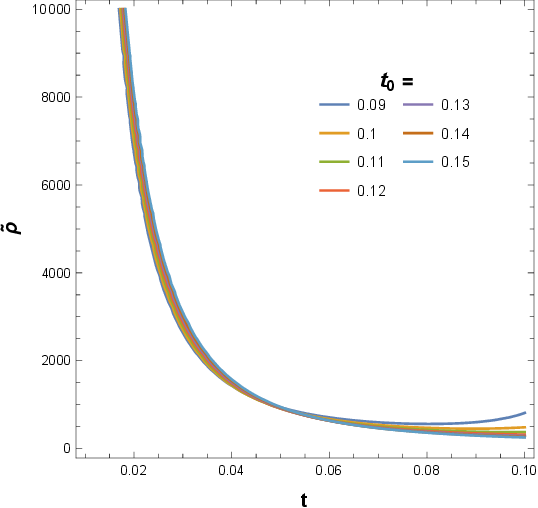}
\caption{Variation of effective energy density $\tilde{\rho}$ with $t$ for $\bar{\o}=-0.9$ for $C'_{1}=1,~C'_{2}=1,~C'_{3}=0.1$}
\label{XIVa}
\end{subfigure}
\end{minipage}
\hspace{0.1cm}
\begin{minipage}[t]{0.3\linewidth}
\centering
 \begin{subfigure}[t]{1.0\textwidth}
        \includegraphics[width=\textwidth]{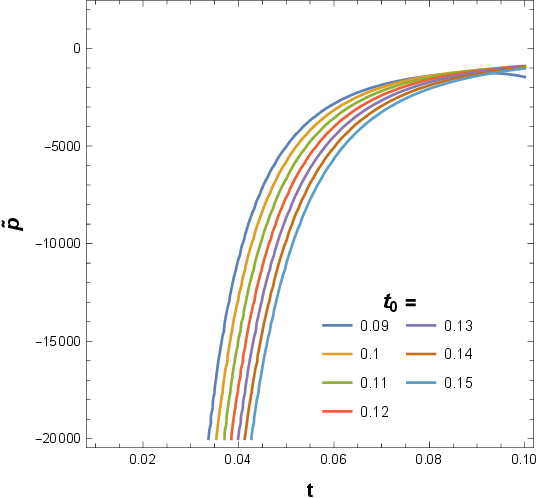}
        \caption{Variation of effective pressure $\tilde{p}$ with $t$ for $\bar{\o}=-0.9$ for $C'_{1}=1,~C'_{2}=1,~C'_{3}=0.1$}
        \label{XIVb}
    \end{subfigure}
\end{minipage}
\hspace{0.1cm}
\begin{minipage}[t]{0.3\linewidth}
\centering
 \begin{subfigure}[t]{1.0\textwidth}
        \includegraphics[width=\textwidth]{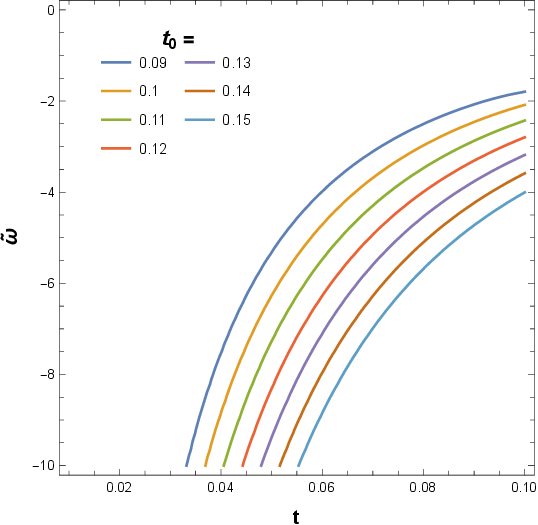}
        \caption{Variation of $\tilde{\o}$ with $t$ for $\bar{\o}=-0.9$ for $C'_{1}=1,~C'_{2}=1,~C'_{3}=0.1$}
        \label{XIVc}
    \end{subfigure}
\end{minipage}
\caption{Variation of effective energy density ($\tilde{\rho}$), effective pressure ($\tilde{p}$) and effective EOS parameter ($\tilde{\o}$) for Case-II}
\label{XIV}
\end{figure*}

\begin{figure*}
\hspace{0.4cm}
\begin{minipage}[b]{0.38\linewidth}
\centering
 \begin{subfigure}[b]{1.0\textwidth}
        \includegraphics[width=\textwidth]{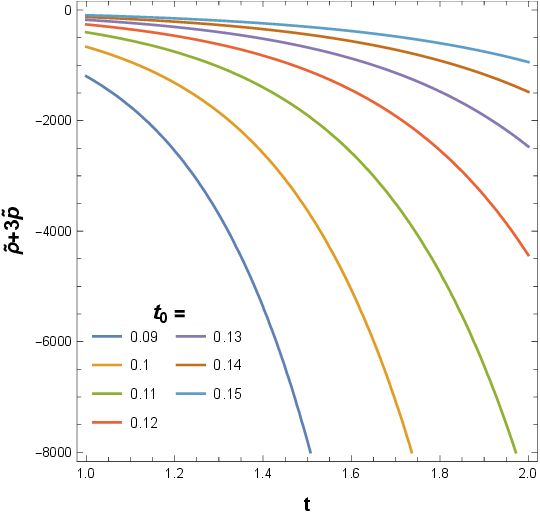}
        \caption{Variation of ($\tilde{\rho}+3\tilde{p}$) with $t$ for $\bar{\omega}=-0.9$ for $C'_{1}=1,~C'_{2}=1,~C'_{3}=0.1$}
        \label{XVa}
    \end{subfigure}
\end{minipage}
\hspace{1cm}
\begin{minipage}[b]{0.38\linewidth}
\centering
 \begin{subfigure}[b]{1.0\textwidth}
        \includegraphics[width=\textwidth]{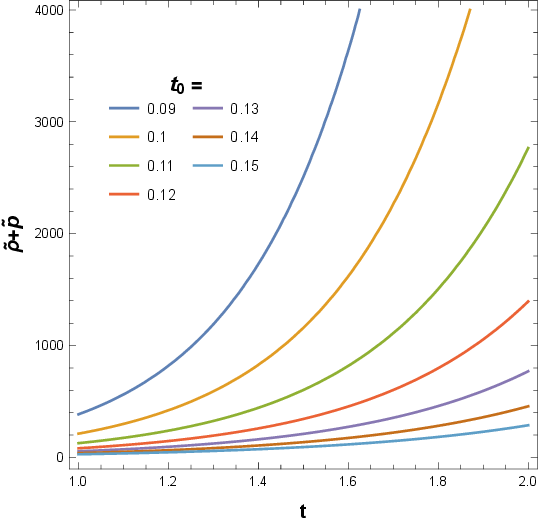}
        \caption{Variation of  ($\tilde{\rho}+\tilde{p}$) with $t$ for $\bar{\o}=-0.9$ for $C'_{1}=1,~C'_{2}=1,~C'_{3}=0.1$}
        \label{XVb}
    \end{subfigure}
\end{minipage}
\caption{($\tilde{\rho}+3\tilde{p}$) and ($\tilde{\rho}+\tilde{p}$) for Case-II}
\label{XV}
\end{figure*}

In the same way, as in the preceding case, we verify for the function (\ref{42}) whether the energy conditions (Eqs. (\ref{48}), (\ref{49}), and (\ref{50})) have been satisfied or not. To verify energy conditions, we utilise Eq. (\ref{42}) in the effective energy density $(\tilde{\rho})$ (\ref{46}) and effective pressure $(\tilde{p})$ (\ref{47}). For this case, we get different plots when we put the function (\ref{42}) in these expressions of $(\tilde{\rho})$, $(\tilde{p})$ and the energy conditions. Figs. (\ref{XIVa}), (\ref{XIVb}) and (\ref{XIVc}) shows the variation of effective energy density ($\tilde{\rho}$), effective pressure ($\tilde{p}$) and effective EOS parameter ($\tilde{\o}$) respectively with time for various $t_{0}$ values. Fig. (\ref{XVa}) shows the variation of $\tilde{\rho}+3\tilde{p}$ and Fig. (\ref{XVb}) represents the variation of $\tilde{\rho}+\tilde{p}$ with time for various $t_{0}$ values.

Fig. (\ref{XIVa}) illustrates the continuous decrease in effective energy density over time for different values of $t_0$, indicating the typical expansion of the cosmos. However, Fig. (\ref{XIVb}) and (\ref{XIVc}) show the negative characteristics of the effective pressure and effective EoS parameters. This indicates that dark energy is dominating the universe in the second scenario of our model. Fig. (\ref{XVa}) illustrates the negative nature of ($\tilde{\rho}+3\tilde{p}$) over time, indicating a violation of the SEC. On the other hand, Fig. (\ref{XVb}) demonstrates the positive nature of ($\tilde{\rho}+\tilde{p}$), implying the satisfaction of the NEC. Furthermore, when considering Figs. (\ref{XIVa}) and (\ref{XVb}), it can be shown that the WEC is also satisfied. The violation of the SEC, as well as the non-violation of the NEC and WEC, imply that the current universe is both expanding and accelerating. Variations in effective pressure and effective EoS, on the other hand, demonstrate that dark energy dominates the cosmos. So, unlike the previous example, we may infer that the second scenario of our model is more relevant with respect to the present observations \cite{Planck1,Planck2,Planck3}, even if the universe's homogeneity and isotropy are doubtful \cite{Marra}.

\section{Conclusion}
In the present study, we analyze both the modified Raychaudhuri equation (\ref{23} or \ref{24}) and the modified field equation of $f(\R,\T)$ gravity (\ref{17}) within the framework of a non-canonical theory known as K-essence.
We use the idea of a perfect fluid that is isotropic and homogeneous (\ref{28}) to combine these two equations to get the Raychaudhuri equation of the K-essence $f(\R,\T)$ gravity (\ref{29}). We also represent the condition (\ref{30}) for which the first Friedmann equation of $f(\R,\T)$ gravity (\ref{19}) is analogous to the Raychaudhuri equation of K-essence $f(\R,\T)$ gravity. However, it is important to note that this article does not discuss the Friedmann equations. Instead, our attention is on the modified Raychaudhuri equation (\ref{29}) of modified $f(\R,\T)$ gravity.

Next, we consider two different processes to solve the differential equation of the modified Raychaudhuri equation (\ref{29}) to get a functional form of $f(\R,\T)$ in the form of Eq. (\ref{34}). It should be pointed out that the additive form of $f(\R,\T)$ is physically and observationally relevant \cite{Harko2,Carvalho,Ordines}. In the first example (Case-I), the scale parameter is considered as a power law of time scale (\ref{32}), which obeys the relation of K-essence EOM (\ref{14}), in obtaining Eq. (\ref{36}). In this process, we took the EOS parameter of K-essence geometry ($\bar{\o}$) to satisfy the EOS value of the accelerated epoch, as in Eq. (\ref{35}).  In the second example (Case-II), we consider the $\K$ to vary with time exponentially to satisfy the criterion ($0<\K<1$) for the well-behaved K-essence FLRW metric (\ref{13}). We then find the dependency of the scale parameter on time (\ref{40}) with the help of the EOM of K-essence (\ref{14}). Then we solve the differential equation of Raychaudhuri in $f(\R,\T)$ gravity (\ref{41}) and establish the form of $f(\R,\T)$ for Case-II (\ref{42}). The form obtained in Case-I and Case-II looks similar, though they are different in the parameters, where they contain as exponentials and coefficients. Hence, the Raychaudhuri equations (\ref{31} and \ref{41}) and the expressions of $f(\R,\T)$ (\ref{36} and \ref{42}) exhibit distinct characteristics when tested under viability analysis and energy conditions.

We graphically examine the exponentials of two cases: $\a$ \text{and} $\gamma$ for Case-I in Fig. (\ref{II}) and $\a'$ \text{and} $\gamma'$ for Case-II in Fig. (\ref{XII}) with $\K$. Then we check the viability of these two cases. For Case-I we plot $f_{\R}(\R,\T)$, $f_{\R\R}(\R,\T)$ and $f_{\T}(\R,\T)$ for a fixed value of $m(=5)$ and varying $\K$ in Fig. (\ref{III}), whereas, in Fig. (\ref{IV}) we plot $f_{\R}(\R,\T)$, $f_{\R\R}(\R,\T)$ and $f_{\T}(\R,\T)$ by keeping $\K(=0.55)$ constant and varying $m$. In Case-II we plot the above said functions taking various values of $\K$ in Fig. (\ref{XIII}). We also plot the effective energy density $(\tilde{\rho})$, effective pressure ($\tilde{p}$) and effective EOS parameter $\tilde{\o}$ for Case-I in Figs. (\ref{IX}), (\ref{X}) and (\ref{XI}) respectively. The same quantities for Case-II have been plotted in Figs. (\ref{XIVa}), (\ref{XIVb}), and (\ref{XIVc}) respectively. Next, to examine the energy conditions for both cases we plot the expressions of ($\tilde{\rho}+\tilde{p}$), ($\tilde{\rho}+3\tilde{p}$) and $\tilde{\rho}$ for Case-I for $\bar{\o}=-0.9$. Figs. (\ref{VII}) and (\ref{VIII}) represents the variation of ($\tilde{\rho}+3\tilde{p}$), ($\tilde{\rho}+\tilde{p}$) and $\tilde{\rho}$ for $\bar{\o}=-0.9$.
Fig. (\ref{XVa}) represents the variation of ($\tilde{\rho}+3\tilde{p}$) for $\bar{\o}=-0.9$ for Case-II. Fig. (\ref{XVb}) represents the variation of ($\tilde{\rho}+\tilde{p}$) for $\bar{\o}=-0.9$ for Case-II.

Now, we discuss the physical significance of our findings in a nutshell for two different cases, as follows:

{\it Case-I}: To assess the feasibility of the first form of the function $f(\R,\T)$ (\ref{36}), we observed that a low value of the parameter $m$ or large values of $\K$ result in a negative value for $f_{\R\R}$. This suggests that the universe is governed by an extreme phase. On the other hand, when we studied the energy conditions through the effective energy density and effective pressure, we learned about the more interesting nature of the universe. The images  (\ref{VII}), and (\ref{VIII}), as well as the table (\ref{TableIV}), show that the SEC and NEC were broken after certain time parameters and $m$ values. This means that the universe is expanding and bouncing in nature. However, based on the information provided in Figs. (\ref{IX}), (\ref{X}), and (\ref{XI}), the model indicates the occurrence of three distinct types of physical events, namely: (i) It is indicated that the universe's development transitions from one area that fulfills the NEC to another area that also fulfills the NEC, while passing through a region where the NEC is violated, similar to prior instances mentioned in Ref. \cite{Arefeva,Arefeva1,Rubakov,Cai,Cai1,Cai2,Singh2023,Ijjas}. This particular form of transition might have observational significance in the examination of gravitational waves \cite{Cai,Cai1,Cai2}. (ii) Secondly, our model demonstrates the presence of the symmetric property during the phase transition in the phase when the NEC is violated. Thus, we may infer that in our model, bouncing events may occur due to a symmetrical violation of the NEC during the expansion of the universe. (iii) Finally, resonant-type quantum tunneling events can occur throughout the time range of $-1$ to $+1$ while considering the behavior of all the specified effective parameters.  The temporal period from $-1$ to $+1$ is traditionally regarded as a classically prohibited region, yet it is permissible within the framework of quantum theory. This phenomenon can be referred to as the resonant-type gravito-quantum effect within a specific range of time scales. The NEC violation has also been investigated in a particular form of scalar field theory, $P(X)$ cosmology. The NEC violation or bouncing scenario entails imposing a shift symmetry or requiring technically unnatural small operator coefficients inside the low-energy effective field theory, thereby carefully tuning the Effective Field Theory (EFT) \cite{Rham}. In our scenario, the K-essence theory is devoid of fine-tuning issues whereas the quintessence model has to need fine-tuning. As a result, the NEC violation and bouncing cosmology occur automatically, without the need for additional coefficients. The NEC violation may be used to examine the universe's late-time acceleration. But for the time being, we will save it for future research. The significance and conclusions of NEC violations have been covered before ({\it vide. Sec.-5.1.3)}).\\

{\it Case II:} In the second version of the function $f(\R,\T)$ (\ref{42}), from Fig. (\ref{XIII}), all the viability criteria (\ref{44}) have been achieved except for large values of $\K$, where $f_{\T}$ is negative. This phenomenon can be attributed to the extensive gravitational interactions with the K-essence scalar field, particularly through the trace of the energy-momentum tensor of this geometry in $f(\R,\T)$ gravity. However, when examining the energy conditions, it is observed from Figs. (\ref{XIV}) and (\ref{XV}) that the effective energy density exhibits a positive but decreasing trend over time. The effective pressure and effective EoS parameters are both negative. The SEC is violated, but the NEC and WEC are fulfilled. This result suggests that the cosmos is undergoing accelerated expansion and is predominantly influenced by dark energy. The examination of the aforementioned figures is observationally significant when considering the isotropy and homogeneity of the cosmos, as indicated by Refs. \cite{Planck1,Planck2,Planck3}. However, the current study demonstrates that the isotropy and homogeneity of the universe are subject to debate \cite{Sarkar, Peebles, Secrest, Marra}.\\

Based on the previous discussions of two cases, we can conclude that by utilizing the Raychaudhuri equation in the framework of $f(\R,\T)$ gravity, which is a non-canonical theory, our model can effectively account for various phenomena in the universe. These include the violation of the NEC, the symmetrical nature of both sides of the NEC violating region, the occurrence of quantum tunneling effects, and the presence of accelerating expansion dominated by dark energy.

As a final comment, we have just focused on the late time acceleration period of the cosmos, neglecting other phases such as the matter-dominated epoch or radiation-dominated epoch. This is because we have examined the power law expansion of the scale factor in the first example and the exponential power law through the exponential variation of $\K$ in the second case. Both cases are related to the accelerated phase of expansion.
Additionally, it is important to mention that this article does not address the study of non-canonical $f(\R,\T)$ gravity in astrophysics, particularly its field equations using the Raychaudhuri equation. Furthermore, no data is used for analysis in this article. However, these topics may be explored in future research.

\vspace{0.2in}
{\bf Acknowledgments:} A.P. and G.M. acknowledge the DSTB, Government of West Bengal, India, for financial support through Grant Nos. 856(Sanc.)/STBT-11012(26)/6/2021-ST SEC dated 3rd November 2023. G.M. expresses gratitude to the authorities of IISER, Kolkata, particularly Prof. N. Banerjee, for the fruitful conversations during the visiting program. All of the authors would also like to thank Prof. S. Ray of GLA University in Mathura, Uttar Pradesh, India and Dr. P. Panchadhyayee of P.K. College in Contai, West Bengal, India, for their helpful conversations. The authors express their gratitude to the referees for providing insightful comments to enhance the manuscript.\\

{\bf Conflicts of interest:} The authors declare no conflicts of interest.\\

{\bf Data availability:} There is no associated data with this article, and as such, no new data was generated or analyzed in support of this research.\\

{\bf Declaration of competing interest:}
The authors declare that they have no known competing financial interests or personal relationships that could have appeared to influence the work reported in this paper.\\

{\bf Declaration of generative AI in scientific writing:} The authors state that they do not support the use of AI tools to analyze and extract insights from data as part of the study process.

\appendix
\section{Expressions of the functions of $\dot{\phi}$}
\ben
g_{1}(\dot{\phi})=\frac{\ddot{\phi}^2(16-49\K)}{12(1-\K)^2}
\label{A1}
\een
\ben
g_{2}(m,\dot{\phi})=\frac{1-\dot{\phi}^4}{2}\frac{1-6m+5m\K}{7m-1-6m\K},
\label{A2}
\een
\ben
g_{3}(m,\dot{\phi})=\frac{(1-\K)\Big[-1+2m+m(3m-4)\K\Big]}{(1-2m+m\K)^2},
\label{A3}
\een
\ben
g_{4}(\dot{\phi})=(1-\K)(1-\sqrt{1-\K}).
\label{A4}
\een

\ben
g_{5}(\dot{\phi})=1-\K
\label{A5}
\een

\ben
g_{6}(\dot{\phi})=a^2(t) (1-\sqrt{1-\K}),
\label{A6}
\een

\ben
g'_{6}=-\frac{\dot{\phi}^2 (16-49 \dot{\phi}^2) (1-\dot{\phi}^2)}{8 (2-7 \dot{\phi}^2)}
\label{A7}
\een 

\ben
g_{7}=a^2(t).
\label{A8}
\een

\ben
h_{1}=\frac{3 (1-\dot{\phi}^4)}{2-7 \dot{\phi}^2}-\frac{36 (1-\dot{\phi}^4)^2}{(2-7 \dot{\phi}^2)^2}
\label{A9}
\een

\ben
h_{2}=\frac{(1-\dot{\phi}^2) (1-\dot{\phi}^4)^2}{6 (2-7 \dot{\phi}^2)}+\frac{1}{6} (\dot{\phi}^4-1)
\label{A10}
\een

\section{Expression of $\a$, $\gamma$, $B$ and $F$}
\subsection{Case-I}
\begin{widetext}
\ben
\alpha &&=\frac{1}{64 ((3 m-4) m \K+2 m-1)^2 (m (6 \K-7)+1)^2}\Bigg[\Big(\Big(m^3 (\K (\K (\K (274 \K-943)+1062)-472)+96)\nonumber\\
&&+m^2 (\K (\K (299 \K-645)+520)-224)+7 m (3 (\K-4) \K+16)-4 (\K+3)\Big)^2\Big)+\frac{2 (m (\K-2)+1)^2}{(3 m-4) m \K+2 m-1}\Bigg]^{1/2}\nonumber\\
&&+\frac{(m (\K-2)+1) (m (m (\K (\K (274 \K-395)+128)-48)+\K (25 \K-8)+32)-4 (\K+1))}{8 ((3 m-4) m \K+2 m-1) (m (6 \K-7)+1)}+1
\label{B1}
\een
\ben
\gamma &&=\frac{1}{8 ((3 m-4) m \K+2 m-1) (m (6 \K-7)+1)}\Big(m^3 (\K (\K (\K (274 \K-943)+1062)-472)+96)\nonumber\\
&&+m^2 (\K (\K (299 \K-645)+520)-224)+7 m (3 (\K-4) \K+16)-4 (\K+3)\Big)~~~
\label{B2}
\een
\end{widetext}
\ben
B=-\frac{5 (\sqrt{1-\K}-1) (m (\K-2)+1)^2}{(3 m-4) m \K+2 m-1}
\label{B3}
\een
\ben
F=\frac{2 (m (\K-2)+1)^2}{(3 m-4) m \K+2 m-1}
\label{B4}
\een

In the absence of the K-essence scalar field ($\dot{\phi}^2=0$), the above parameters are:
\ben
\alpha=&&\frac{1}{98} \Big(-84 m+49 \sqrt{\frac{4 m (m (4 m (9 m+16)-15)-1)+1}{(1-7 m)^2}}\nonumber\\
&&+\frac{5}{1-7 m}+142\Big)
\label{B5}
\een

\ben
\gamma=\frac{96 m^3-224 m^2+112 m-12}{8 (1-7 m) (2 m-1)}
\label{B6}
\een
\ben
B=0,~~~~~~~F=4m-2.
\label{B7}
\een

It is observed that all the aforementioned parameters remain constant in the absence of the K-essence scalar field, since the value of $m$ remains positive and corresponds to the conventional $f(R,T)$ gravity \cite{Panda1}.
\subsection{Case-II}
The expression of $\a'$ for $\o=-0.9$ and $n=\frac{17}{30}$ in Case II

\ben
\a'=&&\frac{1}{24} \Bigg[-\frac{(7 \K-2) \Big(\K (4 \dot{\phi}^6+69 \K+4)-4\Big)}{(\K+1) (3 \K+2) (4 \K-5)}\nonumber\\
&&+\frac{1}{((5-4 \K)^2 (\K+1)^2 (3 \K+2)^2)^{1/2}}\times \nonumber\\
&&\Big(\K (\K (\K (\K (\K (\K (8 \K (\K (14 \K (7 \K-4)\nonumber\\
&&+1373)-2000)+62537)+144028)+159068)-429600)\nonumber\\
&&+26608)+265920)+46144\Big)^{1/2}+24\Bigg]
\label{B8}
\een

\small{
\ben
\gamma='\frac{\K (\K (\K (4 (2-7 \K) \K-195)+230)-372)-248}{24 (\K+1) (3 \K+2) (4 \K-5)}
\label{B9}
\een}
\ben
B'=\frac{5 (1-\dot{\phi}^2) (1-\sqrt{1-\dot{\phi}^2})}{\frac{3 (1-\dot{\phi}^4)}{2-7 \dot{\phi}^2}-\frac{36 (1-\dot{\phi}^4)^2}{(2-7 \dot{\phi}^2)^2}}
\label{B10}
\een
\ben
F'=\frac{2 (1-\dot{\phi}^2)}{\frac{3 (1-\dot{\phi}^4)}{2-7 \dot{\phi}}-\frac{36 (1-\dot{\phi}^4)^2}{(2-7 \dot{\phi}^2)^2}}
\label{B11}
\een

\end{document}